\documentclass[a4paper,fleqn,usenatbib]{mnras}
\usepackage{newtxtext,newtxmath}
\usepackage{float}

\usepackage[T1]{fontenc}
\usepackage{ae,aecompl}

\usepackage{graphicx}	
\usepackage{amsmath}	
\usepackage{amssymb}	

\title[Monte Carlo Simulations of CIRs in Winds]{Monte Carlo Simulations of Polarimetric and Light Variability from Corotating Interaction Regions in Hot Stellar Winds}

\author[Carlos-Leblanc et al.]{
Danny Carlos-Leblanc,$^{1}$\thanks{E-mail: carlos@astro.umontreal.ca}
Nicole St-Louis,$^{1}$
Jon. E. Bjorkman$^{2}$ \&
Richard Ignace$^{3}$
\\
$^{1}$D\'epartement de Physique, Universit\'e de Montr\'eal, C.P. 6128, Succ. Centre-Ville, Montr\'eal, Qu\'ebec H3C 3J7, Canada\\
$^{2}$Ritter Observatory, Department of Physics \& Astronomy, University of Toledo, 2801 W. Bancroft, Toledo, OH 43606-3390, USA\\
$^{3}$Department of Physics \& Astronomy, East Tennessee State University, Johnson City, TN 37614, USA
}

\date{Accepted XXX. Received YYY; in original form ZZZ}

\pubyear{2018}

\begin{document}
\label{firstpage}
\pagerange{\pageref{firstpage}--\pageref{lastpage}}
\maketitle

\begin{abstract}
We use a 3D Monte Carlo radiative transfer code to study the polarimetric and photometric variability from stationary corotating interaction regions (CIR) in the wind of massive stars. Our CIRs are approximated by Archimedean spirals of higher (or lower) density formed in a spherical wind originating from the star and we also made allowance for a bright gaussian spot at the base of the CIR. Comparing results from our code to previous analytical calculations in the optically thin case, we find differences which we attribute mainly to a better estimation of the total unpolarized flux reaching the observer. In the optically thick case, the differences with the analytical calculations are much larger, as multiple scattering introduces extra complexities including occultation effects. The addition of a gaussian spot does not alter the shape of the polarization curve significantly but does create a small excess in polarization. On the other hand, the effect can be larger on the light curve and can become dominant over the resulting CIR, depending on the spot parameters and density of the wind.
\end{abstract}

\begin{keywords}
Polarization - Radiative Transfer - methods: numerical - stars: massive - stars: winds, outflow - stars: Wolf-Rayet
\end{keywords}



\section{Introduction}
\label{sec:Intro}
	Radiation-driven winds are a defining feature of massive stars \citep{Puls2008}. High mass-loss rates cause a significant fraction of their envelope to be lost to the interstellar medium in all phases of their evolution, which combined with their hot and intense radiation flux, contributes to the enrichment, ionization and excitation of the gas and dust in the surrounding medium and ultimately, to the evolution of the stellar populations in galaxies \citep{Langer2012,Kennicutt2012}. It is well known that these outflows are inhomogeneous on a small scale \citep[e.g.][]{Moffat1988}, but spectral line variability has revealed that, in some cases, large-scale structures also form in the winds of these stars. These large-scale structures were first discovered in the solar wind and were later generalized to a broader context of stellar winds \citep[e.g.][]{Mullan1984}. The most common evidence for the presence of these large-scale asymmetries are the discrete absorption components (DACs) that are observed in the absorption troughs of ultraviolet (UV) P Cygni profiles of O stars \citep[e.g.][]{Howarth1989,Kaper1996,Kaper1999}, which have been shown by \cite{Massa2015} to originate at or very close to the stellar surface. However, they are also revealed as large-amplitude variations in the strong optical emission lines of Wolf-Rayet (WR) stars \citep[e.g.][]{Morel1997,Morel1999,Chene2010}. These variations are found to be epoch-dependant, meaning that the periodic changes are generally found at various epochs but their characteristics can evolve. 
	
	Hydrodynamic simulations have shown that a perturbation at the base of an optically thin wind, represented by a bright (or dim) spot on the surface of the star, generates large-scale corotating structures to form out of the interaction between high and low velocity flows as the star rotates, both in the 2D \citep{Cranmer1996} and 3D \citep{Dessart2004} cases. \cite{Brown2004} presented an analytical model to deduce the kinematics of these structures, from optical depth profiles obtained from spectroscopic observations. These large spiral-like structures have been appropriately named "Corotating Interaction regions" or "CIRs". The epoch-dependant nature of the above-mentioned periodic spectroscopic changes could then be attributed to the dissipation and regeneration of the CIRs. 
	
	A major effort to study CIRs came from the \textit{IUE} MEGA campaign \citep{Massa1995} in which different types of massive stars were monitored in UV spectroscopy; WN5 \citep{St-Louis1995}, B0.5 Ib \citep{Prinja1995} and $\zeta$ Puppis \citep{Howarth1995}. More recent efforts include characterizing the wind structure of WR1 using spectropolarization \citep{St-Louis2013} as well as a detailed study of $\zeta$ Puppis from extensive time-dependant photometry and spectroscopy \citep{Ramiaramanantsoa2018}.
	 
	In two previous publications, we have presented a simplified analytical model for polarimetric variability from such CIRs by describing them parametrically as a spiral-like density enhancement in an otherwise unperturbed spherical wind. In \cite{Ignace2015}, we developed the model, expanding on \cite{Ignace2009}, in the optically thin electron scattering limit and allowing multiple CIRs to be placed on the star, at arbitrary latitudes and azimuth. Our model polarization curves present clear phase-dependant signatures for one or two CIRs, but for multiple CIRs create more complex behaviours. In \cite{St-Louis2018} we extend our model to optically thick winds by accounting for multiple scattering using a "core-halo" approach. This approach defines a pseudo-photosphere beyond the radius of the star, above which the wind can be treated as optically thin. This way, we treated this pseudo-photosphere as the source from which the light emerges, while still having treated the wind and the CIR as initiating from the actual stellar radius. That model was then applied to polarimetric observations obtained from the literature of the WR star WR6, well known to show consistent periodic ($P = 3.77$ days) but epoch-dependant photometric, polarimetric and spectroscopic variability without convincing evidence for the presence of a companion. A Levenberg-Marquardt (LM) nonlinear least-squares minimization algorithm was developed to fit 13 different datasets obtained over a time span of about five years (as well as two older datasets from twenty years before). Two CIRs were used, and a number of parameters related to the stellar wind were adopted. The algorithm was able to fit all observations with consistent stellar parameters and found a stellar inclination of $166^\circ$ and an orientation of the stellar axis on the plane of the sky of $63^\circ$. In all cases, the CIRs were found to be located close to the stellar equator and separated by approximately $90^\circ$ in longitude. Only their specific locations on the stellar surface were found to differ from one epoch to the next.
	
	In this paper, we expand upon these two papers using Monte Carlo radiative transfer (MCRT). Section \ref{sec:Desc} describes the MCRT model, and Section \ref{sec:val} presents validation tests to determine the limit between the optically thin and optically thick cases, as well as the number of photons required to obtain significant results. We also present error estimates for different wind densities introduced by varying the random number generator seed. In Section \ref{sec:comp} we compare our polarimetric calculations to those obtained both for the optically thin \citep{Ignace2015} and thick \citep{St-Louis2018} limits. Finally, in Section \ref{sec:GauSpot} we present a parameter study for polarization and light curve calculations including a CIR and an associated bright gaussian spot. For these calculations, we have used stellar parameters typical of Wolf-Rayet stars. We conclude in Section \ref{sec:conc}.

\section{The Monte Carlo Radiative Transfer (MCRT) Model}
\label{sec:Desc}

	The principle behind MCRT simulations is to follow a large number of monochromatic energy packets, each containing an ensemble of individual photons. So the packet, hereafter referred to as a "photon" may be partially polarized. These packets are randomly emitted from the stellar surface and travel through some medium, in our case a hot stellar wind until they escape the system. Collectively, these photons represent the luminosity emitted by the star. The scattering during radiation transport is determined by randomly sampling the optical depth while the change in direction (and polarization) of the packet is determined by randomly sampling the Rayleigh phase function. Photons are followed until they leave the envelope where they are placed into the appropriate latitude, azimuth and frequency bins. To improve our signal-to-noise ratio, we have also implemented a source function sampling procedure. This procedure samples every photon interaction, which measures the scattering source function, and then emits virtual photons in the direction of every individual viewpoint. The result is then weighted by the probability of the virtual photon to scatter and subsequently escape in that direction \citep[often called photon peeling, e.g.][]{Yusef-Zadeh1984,Whitney2011}. 
	
	Our goal with these Monte Carlo simulations was to study the polarimetric and photometric changes caused by the presence of a density perturbation representing the CIRs. We employed a three-dimensional approach for the wind model by adopting a time-independent spherical wind from a hot star, threaded by a CIR as a density perturbation in the shape of a 3D Archimedean spiral. This simplified structure is not quite equivalent to that predicted by hydrodynamic calculations \citep[e.g.][]{Cranmer1996} but our goal with these initial numerical models is to compare our results with our analytical calculations. In future work, we intend to complexify the density and velocity structures of our CIRs. Rotation was simulated by having observers view the star from different azimuths along the same "inclination" or latitude, which act as different rotation phases for the star. As electron scattering is the dominant source of polarization in hot stellar winds, we neglected all other types of scattering. Since electron scattering is wavelength independent, we used a monochromatic approach for the purposes of this paper. 

\subsection{Polarization}
\label{subsec:pol}

	We used Stokes vectors to describe the polarization in our simulation, as described by \cite{Chandrasekhar1960}. Stokes vectors are separated into the four parameters $I$, $Q$, $U$ and $V$ by \citep[e.g.][]{Whitney2011} : 	
	\begin{equation}\label{eq:Stokes}
		\bold{S}(\theta ,\phi) = [I(\theta ,\phi), Q(\theta ,\phi), U(\theta,\phi), V(\theta ,\phi)],
	\end{equation}
	
\noindent where $I$ describes the intensity of the incoming light, $Q$ and $U$, the linear polarization components and $V$, the circular polarization. These components depend on the spherical coordinates $\theta$ and $\phi$ in the star's frame of reference, and therefore obviously differ depending on the observer's frame of reference. In our model, $V(\theta,\phi)$ is always 0 as we assume that there are no sources of circular polarization in the wind. Photons emitted by the star have an initial Stokes vector of $(1,0,0,0)$ and travel a distance characterized by a random optical depth of $\tau = -\ln(1-\epsilon)$, where $\epsilon$ is a random number between 0 and 1, which is chosen from a Poisson distribution with unit mean. After travelling this distance, they scatter and are given a new direction $(\theta',\phi')$ using the Mueller and phase function matrices which rotate the frame of reference from the previous scattering frame to the next \citep[e.g.][]{Code1995}. The resulting Stokes vector is expressed as
	\begin{equation}\label{eq:NewStokes}
		\bold{S} = \bold{R}(\chi)\bold{L}(i)\bold{S'},
	\end{equation}
	where $\bold{S'}$ is the previous Stokes vector. $\bold{L}(i)$ is the Mueller rotation matrix for an angle $i$ 
	\begin{equation}\label{eq:Mueller}
		\bold{L}(i) = 
		\begin{bmatrix}

			1 &      0 &     0 & 0  \\
			0 &  \cos2i & \sin2i & 0  \\
			0 & -\sin2i & \cos2i & 0  \\
			0 &      0 &     0 & 1
		\end{bmatrix}
	\end{equation}
	and $\bold{R}$ is the Rayleigh phase function matrix for a scattering angle $\chi$, given by 
	\begin{equation}\label{eq:Rayleigh}
		R(\chi) = 
		\begin{bmatrix}

			\cos^2\chi +1 & \cos^2\chi -1 & 0        & 0        \\
			\cos^2\chi -1 & \cos^2\chi +1 & 0        & 0        \\
			           0 &            0 & 2\cos\chi & 0        \\
			           0 &            0 &        0 & 2\cos\chi          
		\end{bmatrix}.
	\end{equation}

	Once the photon has escaped from the wind, the Stokes vector reference direction is realigned to the stellar rotation axis, from which we binned the photons into the appropriate observer bins depending on the inclination. Once the simulation is finished, the final Stokes vector is normalized by the total flux for each bin so that
	\begin{align}\label{eq:StNorm}
	\begin{split}
	 q = Q/I ,
	\\
	 u = U/I .
	\end{split}
	\end{align}	
	We then correct for the rotation angle of the stellar axis on the plane of the sky $\psi$ by rotating the final Stokes vectors using $\bold{L}(\psi$).

\subsection{Wind and Grid Properties}
\label{subsec:grid} 
	Our spherical wind was described by a three-dimensional adaptive mesh spherical grid in ($r$, $\theta$, $\phi$), with each cell containing information on the density within it. The density depends on a number of input parameters, as well as on the cell's distance from the star and whether or not a CIR crosses its path. Following our analytical approach, the density of the wind itself is given by 
	\begin{equation}\label{eq:rho_wind}
		\rho_{\rm wind} = \frac{n_0 \mu_e m_H}{w \tilde{r}^2},
	\end{equation}
	where $n_0$ is the number density scaling factor given by
	
	\begin{equation}\label{eq:n0}
		n_0 = \frac{\dot{M} / \mu_e m_H}{4 \pi R_*^2 v_{\infty}},
	\end{equation}
$\mu_e$, the mean molecular weight per free electron, $m_H$, the mass of a hydrogen atom, $\tilde{r}$, the radial distance from the center of the star normalized to the stellar hydrostatic radius $R_*$, and $w(\tilde{r})$ is the normalized velocity of the wind at a location $\tilde{r}$ given by
	\begin{equation}\label{eq:omega_wind}
		w(\tilde{r}) = \frac{v(\tilde{r})}{v_\infty} = 1 - \frac{(1 - w_0)}{\tilde{r}},
	\end{equation}
	where $w_0 = v_0/v_\infty$, with $v_0$ being the velocity at the base of the wind and $v_\infty$, the terminal velocity of the wind. Note that for simplicity, we adopted a standard beta velocity law with an exponent of 1, which is why this exponent does not appear explicitly in the above equation. The number density scaling factor is related to the optical depth parameter $\tau_0$ by 
	\begin{equation}\label{eq:tau0}
		\tau_0 = n_0 R_* \sigma_T,
	\end{equation}
	where $\sigma_T$ is the Thomson scattering cross-section.
	We defined grid cells to become radially larger further out in the wind due to the fact that the density varies as $1/r^2$. However, we have added an extra layer of fine structure for the grid cells containing the CIR in the form of sub-grids for these particular cells. That way, we can resolve the CIR with a much better precision. For our simulations, the coarse grid contains 43 $r$ cells (going from $R_*$ to $100R_*$), 18 $\theta$ cells and 24 $\phi$ cells. The fine grid for the cells containing a sub-grid have an extra set of ($r$,$\theta$,$\phi$) cells depending on the CIR half-opening angle $\beta_0$ and the winding rate of the CIR, which is characterized by the ratio of the rotation and wind terminal velocities, $v_{\rm rot}/v_\infty$. The star itself contains 1 $r$ cell and the same number of $\theta$ and $\phi$ cells as the wind.
	
	Each CIR that crosses the center point of a grid cell increments it's density. The density in a given cell in our grid is therefore given by
	\begin{equation}\label{eq:rho_CIR}
		\rho_{\rm cell} = \rho_{\rm wind}\left(1 + \sum_{\rm CIRs} \eta\right) ,
	\end{equation}
	where $\eta$ is the density contrast with respect to the spherical wind in a CIR crossing the center point of the cell given by 
	\begin{equation}\label{eq:eta}	
		\eta = \frac{n_{\rm CIR}-n_{\rm wind}}{n_{\rm wind}},
	\end{equation}
	with $n_{\rm CIR}$ the number density in the CIR and $n_{\rm wind}$ that of the spherical wind. $\eta=0$ for a spherical wind. Note that we have adopted the approach, described in \cite{Ignace2015}, in which the CIR shape is determined from the radial streamline flow in the rotating frame of reference of the star, corresponding to a spiral pattern as seen by an observer. The center position $\phi_s$ of this spiral is defined by 
	\begin{equation}\label{eq:phi_s}
		\phi_s = \phi_0 - \frac{v_{\rm rot}}{v_\infty}\sin(\theta_{\rm CIR})\left[\tilde{r}-1 + (1-w_0)\ln\left(\frac{\tilde{r}-1+w_0}{w_0}\right)\right],
	\end{equation}
	where $\phi_0$ is the azimuth of the CIR at the base of the wind, $\theta_{\rm CIR}$ is the latitude of the CIR and $v_{\rm rot}$ is the equatorial rotation speed of the star. 

Note that we characterize the amount of winding of our CIRs with the winding radius defined by 
	\begin{equation}
	\label{eq:Rpitch}
		r_0 = \frac{v_\infty}{2\pi/P} = \frac{v_\infty}{v_{\rm rot}}R_*,
	\end{equation}
$P$ being the period of rotation. This parametrization defines the degree of curvature of the spiral. Higher values of $v_\infty/v_{\rm rot}$ correspond to less curved CIRs.
	
\subsection{Spot model}
\label{subsec:spot}
	Our model has the option to add gaussian spots at the base of the CIRs. The luminosity contribution of a spot at a given point on the surface of our star is given by
	\begin{equation}\label{eq:L_spot}
		L_{\rm spot} = Ae^{-\beta / \sigma^2},
	\end{equation}
	where 
	\begin{equation}
		\beta = \cos^{-1}\left(\hat{n} \cdot \hat{n}_{\rm CIR}\right),
	\end{equation}
	$A$ is the intensity of the spot relative to the star, $\hat{n}$ is the unit surface vector at a given location on the star, $\hat{n}_{\rm CIR}$ is that of the center of the base of the CIR related to the spot and 
	\begin{equation}
		\sigma = \beta_0 / (\ln2)^{1/2}
	\end{equation} 
	with $\beta_0$ the half width at half-maximum of the spot, which also corresponds to the half-opening angle of the CIR. 
	The luminosity of a given point on the surface of the star is then given by the sum of the contributions from each spot plus the contribution from the base photosphere. This gives a luminosity map of the whole star, which in turn provides the probability per surface area for a photon to be emitted at this position, normalized by the total luminosity.

\subsection{Input Parameters}
	Our model requires several input parameters to describe both the spherical wind and the CIRs that we define below. 
	First, there are a number of stellar and wind parameters :
	\begin{itemize}
	\item $N_{\rm phot}$, the number of photons in the Monte Carlo simulation,
	\item $R_*$, the stellar radius,
	\item $v_{\rm rot}$, the equatorial rotation rate,
	\item $\tau_0$, the optical depth scaling factor (see equation \ref{eq:tau0}),
	\item $v_0$, the speed of the wind at its base,
	\item $v_{\infty}$, the terminal velocity of the wind.
	\end{itemize}
	There are also a number of CIR input parameters :
	\begin{itemize}
	\item $N_{\rm CIR}$, the number of CIRs,
	\item $\phi_0$, the azimuth of a given CIR,
	\item $\theta_{\rm CIR}$, the colatitude of a given CIR,
	\item $\beta_0$, the CIR half-opening angle,
	\item $\eta$, the density contrast of the CIR (see equation \ref{eq:eta}),
	\item $L_{\rm spot}/L_{\rm phot}$, the ratio of the luminosity of the spot at the base of the CIR to the stellar luminosity.
	\end{itemize}

	We also require a number of parameters associated with the different viewpoints, such as their total number, the inclination and azimuth relative to the star of each individual viewpoint. To simulate rotation, we take data from a series of viewpoints at the same inclination but different azimuths. These act as different rotation phases of the star.
	
	Unless stated otherwise we have adopted these parameters that are thought to be appropriate for the star WR6:
	\begin{itemize}
	\item $R_* = 2.65R_\odot$ \citep{Hamann2006},
	\item $v_0 = 57$ km/s (obtained by assuming $v(r) = v_\infty\left(1-\frac{bR_*}{r}\right)$ with $b=0.97$),
	\item $v_\infty = 1900$ km/s \cite[see][]{St-Louis1995}.
	\end{itemize}

	As for default CIR parameters, unless it is explicitly mentioned, CIRs always have :
	\begin{itemize}
	\item $\eta = 1$,
	\item $\theta_{\rm CIR} = 90^\circ$ and $\phi_0 = 0^\circ$,
	\item $\beta_0 = 15^\circ$.
	\end{itemize}
	Also note that $L_{\rm spot}/L_{\rm phot} = 1$, unless explicitly noted that spots were used in the model.
	
	One last input parameter we must set is the seed, a number which initializes a sequence of (pseudo) random numbers that the MCRT code uses. This is useful to reproduce simulations for debugging purposes, but it can also be useful to vary for estimating statistical error, which is what we do in Section \ref{ssec:err}.
\section{Model Validation}
\label{sec:val}
	To verify the validity of our model, we have performed tests described in this section. First we determine the transition between an optically thin and thick wind. Second we explore how polarization values change as a function of the total number of photons. Finally we characterize the effect of the chosen seed for our MCRT simulations.
\subsection{Optical Thickness}
\label{ssec:opthick}
	To demarcate thin and thick limits, we carried out a series of linearity tests to determine how $q$ and $u$ behave as $\tau_0$ increases. The linear polarization for various values of a CIR's "winding" radius, $r_0$, defined in equation \ref{eq:Rpitch} were calculated for different inclinations of the stellar rotation axis. Our assumption is that a departure from linearity indicates that multiple scattering effects due to optical depth start to become significant.
	\begin{figure}
		\includegraphics[width=\columnwidth]{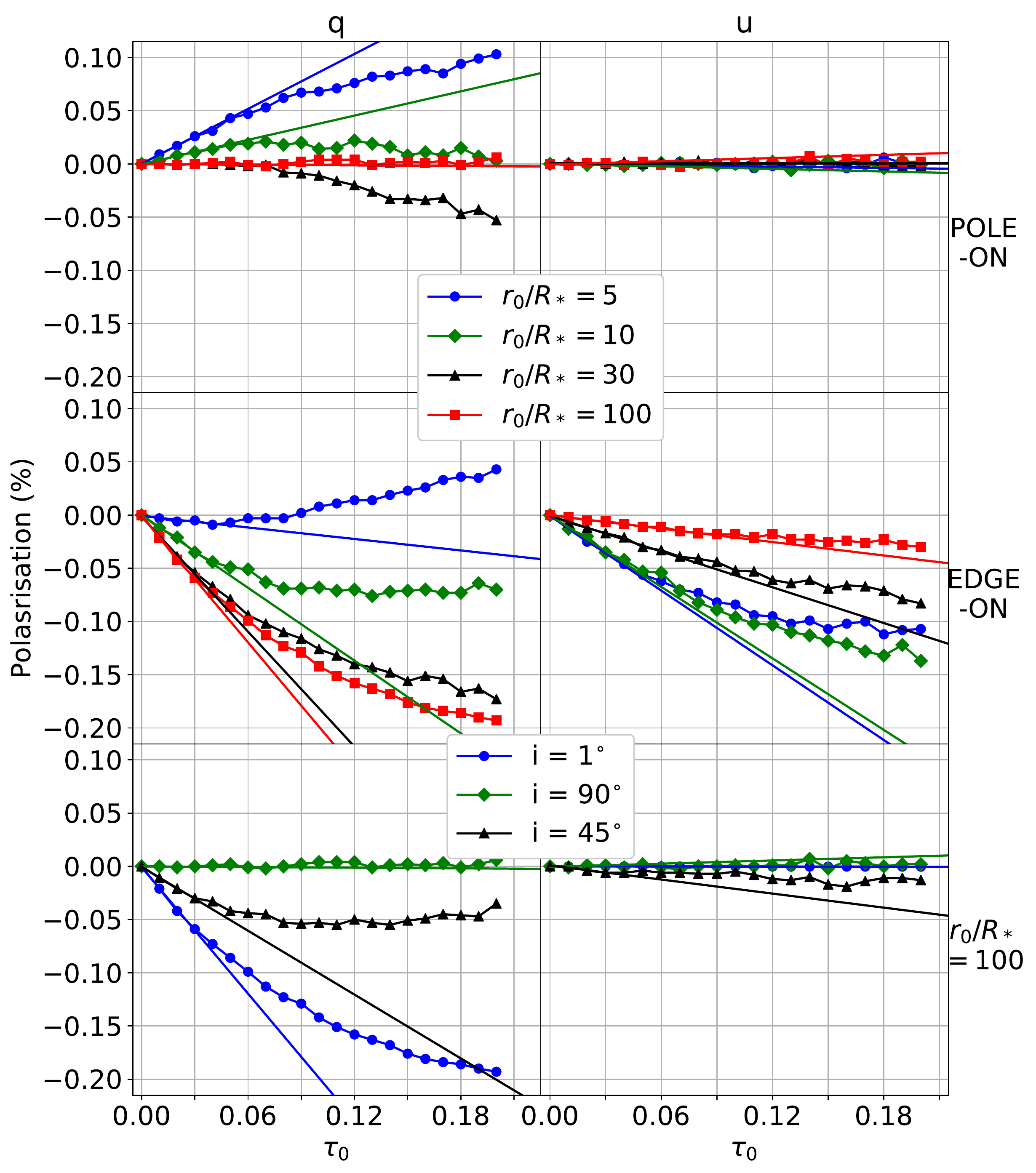}
 		\caption{Linear polarization as a function of $\tau_0$ for an edge-on (top row) and a pole-on view (middle row) for one CIR located in the stellar equator for various values of the winding radius from $5$ to $100R_*$. In the bottom row, we present the linear polarization for an essentially straight CIR ($r_0/R_* = 100$) from different viewing angles. The solid lines represent a linear fit of the first few values showing up to what point the change in polarization remains linear.}
 		\label{fig:Qshift}
	\end{figure}
	For these tests, we calculated only the linear polarization for a single viewpoint at $(\theta,\phi) = (0,0)$ and determined how it varies with $\tau_0$ and $r_0/R_*$. In Figure \ref{fig:Qshift} we present the calculated $q$ and $u$ values as filled circles for the edge-on (top) and pole-on (middle) views when the winding radius is varied from $r_0/R_* = 5$ to $r_0/R_* = 100$, the latter corresponding to an essentially straight CIR. For technical reasons (the reference direction of the observed Stokes vectors is the projection of the stellar rotation axis on the plane of the sky, which is undefined for $i =$ 0), we cannot strictly use $i = 0^\circ$ so our pole-on calculations are in reality for $\theta = 1^\circ$. In the bottom panels, we present the polarization for $r_0/r_* = 100$ as viewed from different inclinations. To determine where the linear polarization starts to deviate for a linear behaviour, we fitted a straight line to the first few values at small $\tau_0$. The fits appear as solid lines. From this figure we conclude that departures from a linear behaviour begin roughly around $\tau_0 = 0.03$,  although for some of the curves a non-linear behaviour only begins for higher values of $\tau_0$. Note that the total electron-scattering optical depth integrated along the line-of-sight to the observer is given by
	
	\begin{equation}
		\tau _e = \int^{\infty}_{R_*}n_e(r)\sigma _T dr	,	
	\end{equation}
	with $n_e(r)=\rho _{wind}(r)/\mu _e m_H$.  Using equation \ref{eq:rho_wind}, this can be expressed as
	
	\begin{equation}
		\tau _e = n_0\sigma _TR_*^2 \int^{\infty}_{R_*}\frac{1}{w r^2} dr=\tau _0R_*\int^{\infty}_{R_*}\frac{1}{w r^2} dr,
	\end{equation}
which results in 
	
	\begin{equation}
		\tau _e = \frac{\tau _0}{b} \ln \left ( \frac{1}{1-b}\right).
	\end{equation}
For our adopted value of $b=0.97$, this yields $\tau_e ={\rm 3.6\ } \tau _0$.  In order to encompass as much as possible all different winding radii and inclinations, we have adopted $\tau_0 =  0.03$ as our optically thin limit; anything higher will be considered as an optically thick calculation. This limits corresponds to $\tau _e$ of the order of 0.1, a value much smaller than the value usually considered to be the optically thick limit, i.e. $\tau _e\approx 1.$ Note that the above results are for a density contrast in the CIR of $\eta = 1$.
\subsection{Photon Numbers}
\label{ssec:nphot}
\begin{figure}
		\includegraphics[width=\columnwidth]{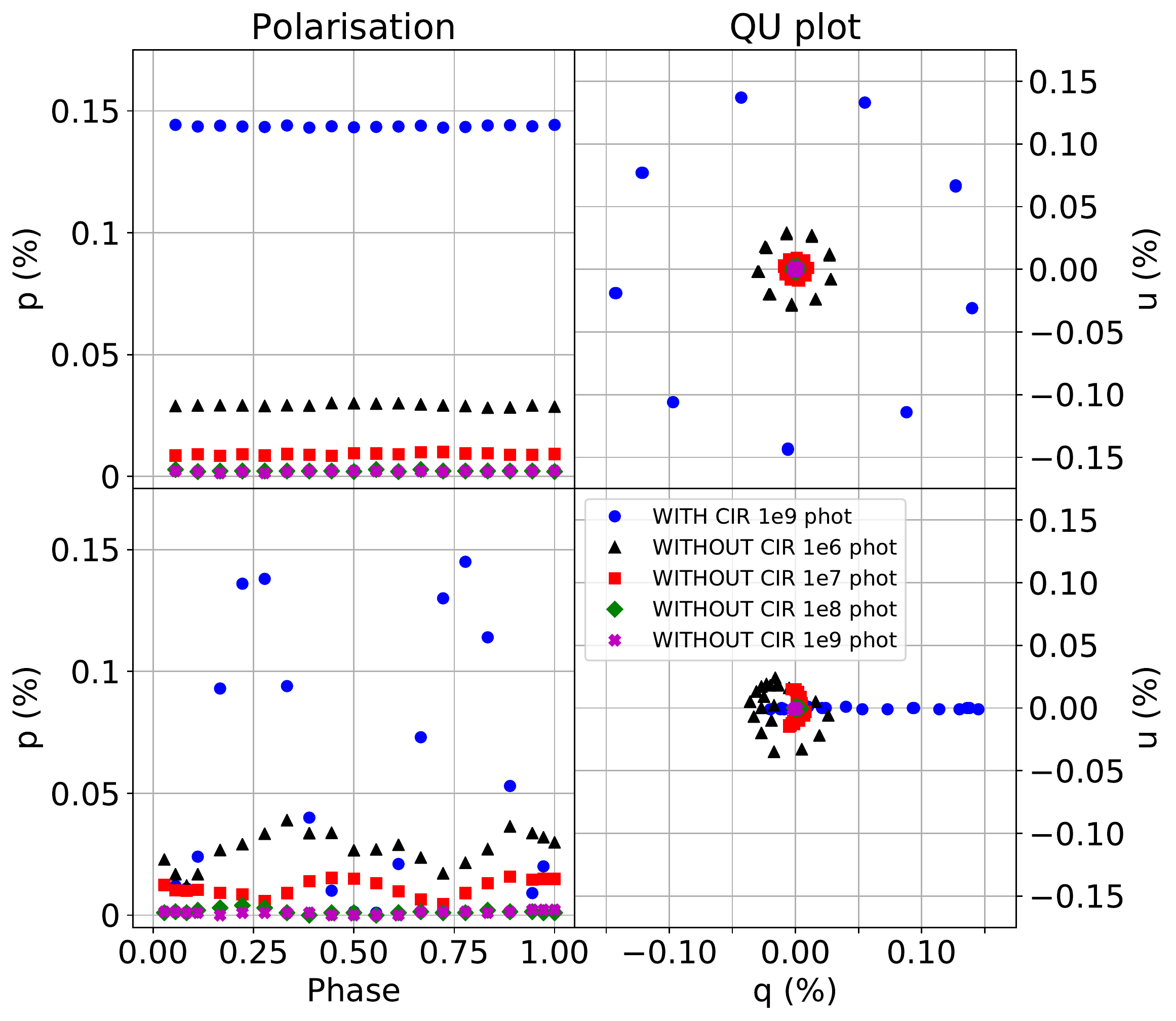}
 		\caption{Polarization contribution from a spherical wind for runs with different numbers of photons in our simulation. We have also included our calculations of the polarization from a spherical wind with a CIR with $\tau_0 = 0.1$ for comparison.}
 		\label{fig:NoCIR}
	\end{figure}

	For a spherical wind, the polarization should be zero. In Figure \ref{fig:NoCIR}, we present our calculated polarization as a function of phase (left panels) and in the $q-u$ plane (right panels) for a spherical wind only, with an intermediate optical depth of $\tau_0 = 0.1$ for a pole-on (top) and edge-on (bottom) view using different numbers of photons for each run, from $1 \times 10^6$ photons to $1 \times 10^9$ photons. For comparison, we also present the polarization from a CIR at the equator with $r_0 = 100R_*$ for both viewing angles when using the highest number of photons.
	
	To minimize the random noise while minimizing the numerical error associated with the spherical wind, as well as the computation time, the number of photons used in the simulation needs to be optimized. From Figure \ref{fig:NoCIR}, with only a million photons, a residual polarization for our spherical wind of about $0.03\%$ is obtained. As the number of photons is increased, the residual polarization gradually decreases. Finally, very little difference is apparent between simulations with $N_{\rm phot} = 1 \times 10^8$ and $N_{\rm phot} = 1 \times 10^9$ for both the pole-on and edge-on views. We therefore decided to use $N_{\rm phot} = 1 \times 10^8$ for all simulations presented in this work. We also notice qualitative differences in the shape of the pole-on and edge-on q-u noise curves. Indeed, a wind that is viewed nearly pole-on leads to a circular pattern in this q-u plane, while a wind viewed edge-on does not show such a clear pattern. Instead, the polarization values seem to be aligned along a preferred axis in the q-u plane. This behaviour is readily explained by our source function sampling algorithm. Each emitted photon and photon interaction emits a virtual photon in the direction of each viewpoint weighted by the probability it has to scatter towards them. This results in a correlation between viewpoints due to each of them receiving the same virtual photons (more precisely, source function sample events), which are simply weighted differently. For this particular case ($\tau = 0.1$) the residual polarization of the spherical wind for $1 \times 10^8$ photons is of the order of $0.003\%$ for both pole-on and edge-on views, which is essentially negligible compared to that of a CIR.
	
\subsection{Statistical Error from Seed Values}
\label{ssec:err}
\begin{figure}
		\includegraphics[width=\columnwidth]{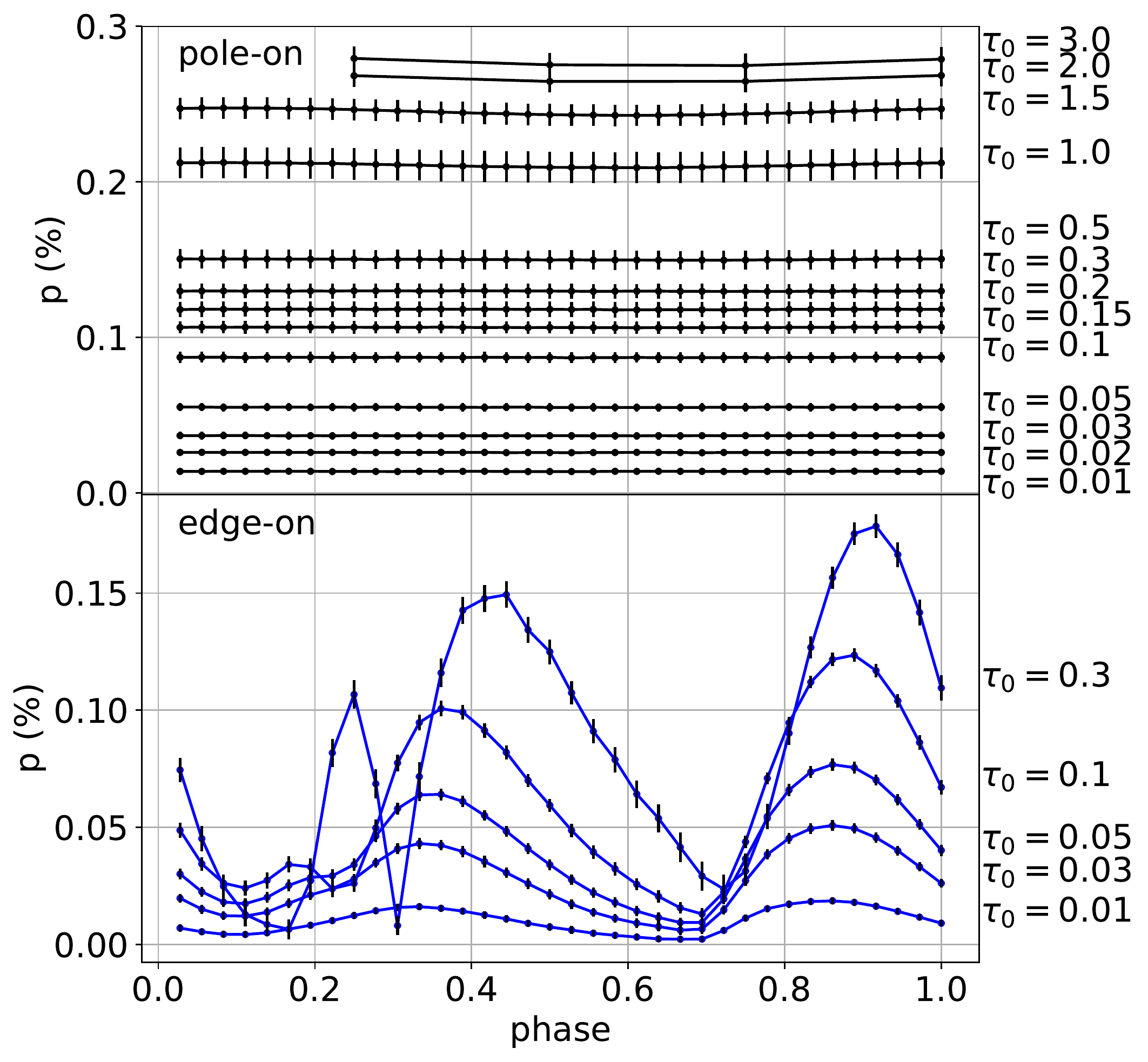}
 		\caption{Pole-on and edge-on polarization values as a function of phase for a CIR with $r_0 = 5R_*$ located at the equator for different values of $\tau_0$ from $0.01$ to $3.0$. The error bars represent the standard deviation given by running the same simulation with 20 different seeds.}
 		\label{fig:diffSeed}
	\end{figure}

	To determine an approximate numerical error bar for the calculated values from our simulations, we performed a series of simulations with identical input parameters, but for different initial seeds for both the pole-on and edge-on view for one equatorial CIR with a winding radius of $r_0 = 5R_*$ for different values of $\tau_0$. For each $\tau_0$ value, we performed 20 simulations and calculated the mean polarization and the standard deviation. Figure \ref{fig:diffSeed} shows our calculated mean polarization values and associated standard deviations as a function of phase for different values of $\tau_0$. In the top panel we show a pole-on view for $\tau_0$ from 0.01 to 3 while in the bottom panel we show an edge-on view from $\tau_0 = 0.01$ to $0.3$. Note that as computations with large values of $\tau_0$ are very expensive in computing time. Therefore, for those cases, we only calculated the polarization values for a small number of phases for the pole-on view. As expected, both the polarization and its the standard deviation generally increases with the number of scatterings, characterized by larger values of $\tau_0$. 
	
	\begin{figure}
		\includegraphics[width=\columnwidth]{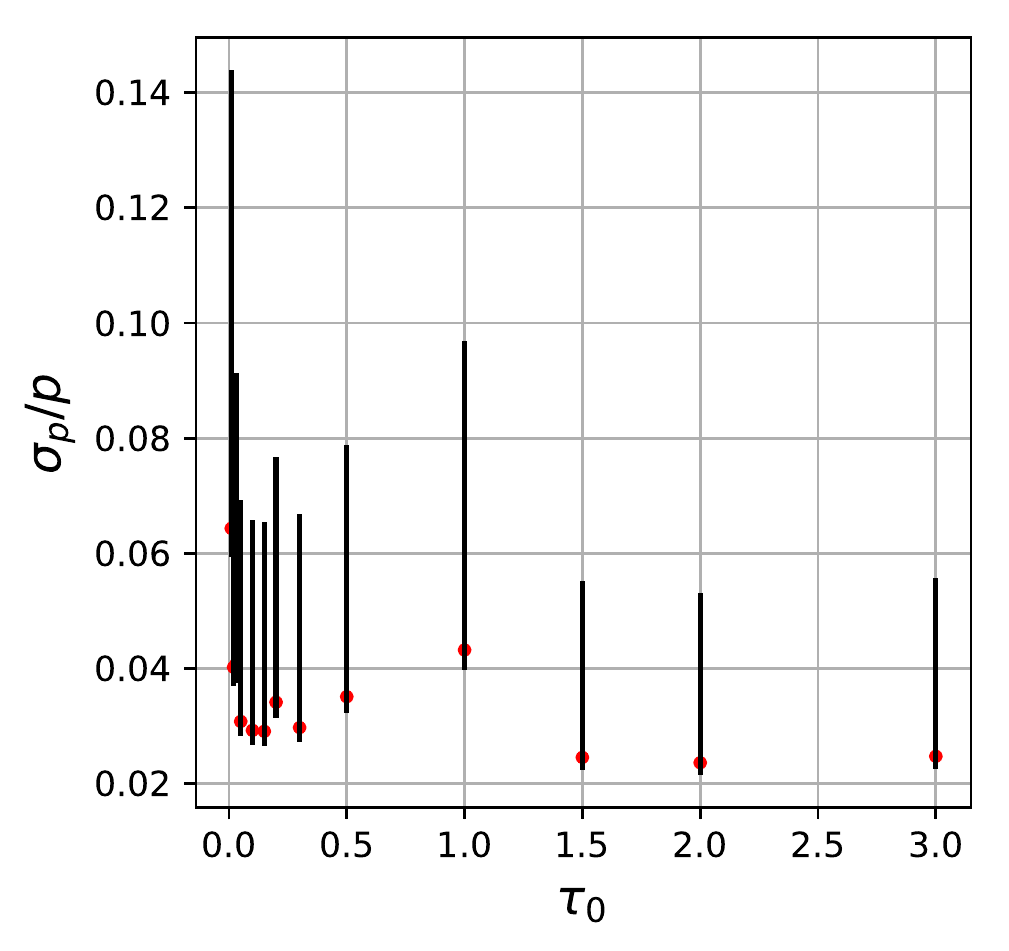}
 		\caption{Mean relative errors a a function of $\tau_0$ for the different curves in the top graph of Figure \ref{fig:diffSeed}.}
 		\label{fig:errseed}
	\end{figure}
 	
 	\begin{figure}
		\includegraphics[width=\columnwidth]{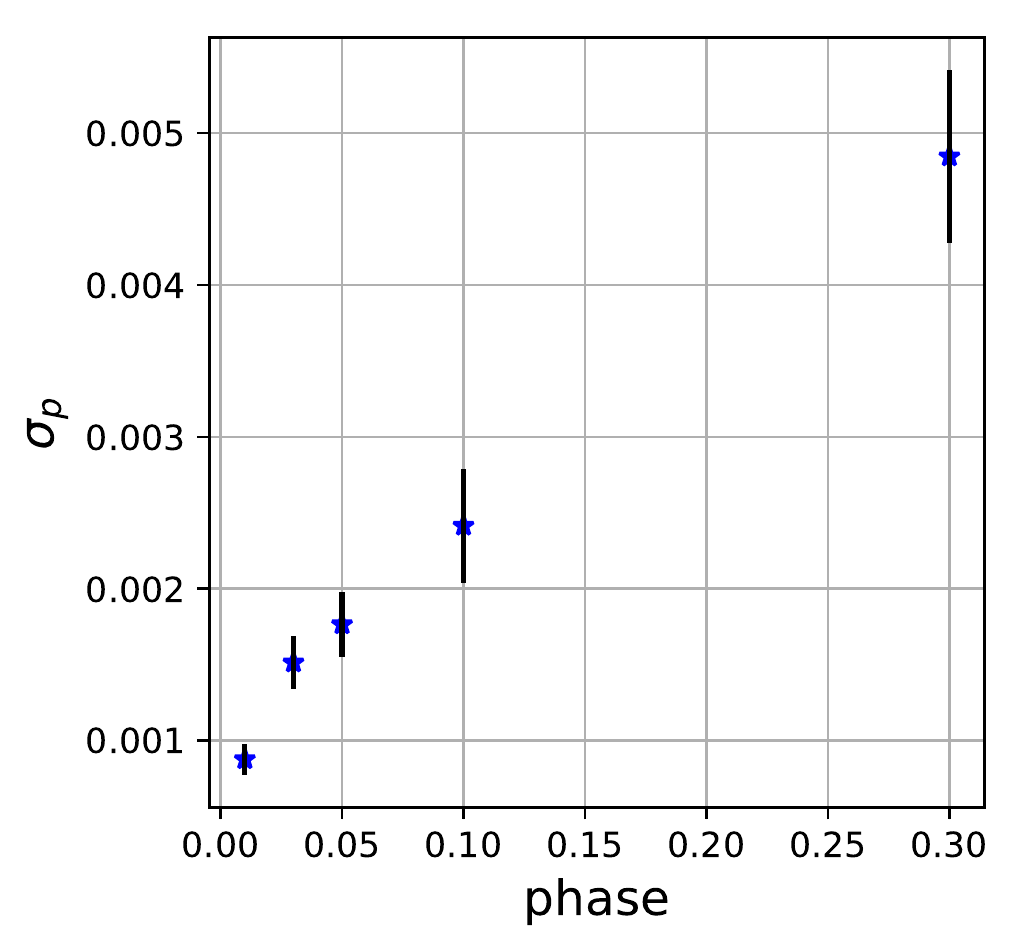}
 		\caption{Mean of the standard deviations for the different curves in the bottom graph of Figure \ref{fig:diffSeed} as a function of $\tau_0$.}
 		\label{fig:errseed2}
	\end{figure}
	To show the behaviour between the standard deviation, $\sigma _p$, and $\tau_0$ we use the pole-on case for which the polarization is nearly constant with phase. In Figure \ref{fig:errseed}, we plot the relative error, defined by the mean of the ratio between the standard deviation and the polarization at each phase as a function of $\tau_0$ for the pole-on view. The vertical error bars in this plot were calculated using error propagation.  The error on the mean value of the polarization, $P$, was set to $\sigma _p/N$, where $N$ is the number of simulations (in this case, $N=20$), while the error on $\sigma _p$ was set by calculating the confidence interval for the variance at the 95\%\ level, assuming a normal distribution.  The latter uncertainty dominates over the error on the mean of the polarization leading to asymmetric error bars.  The behaviour of the relative error as a function of $\tau_0$ can be explained roughly as follows. Assuming that the relative error can be drawn from a Poisson distribution, it should be inversely proportional to the square root of the number of photons that scatter at least once given by 
	\begin{equation}
		\frac{\delta p}{p} \propto \frac{1}{\sqrt{N_{\rm phot}(1-e^{-\tau_0})}} .
	\end{equation}
If we carry out N such calculations, the standard deviation of the mean should be proportional to $1/\sqrt{N}$ and therefore the measured error should vary as
	\begin{equation}
		\frac{\delta p}{p} \propto \frac{1}{\sqrt{N \cdot N_{\rm phot}(1-e^{-\tau_0})}} .
	\end{equation}
At small $\tau_0$, $\delta p / p$ should therefore be proportional to $\frac{1}{N \cdot N_{\rm phot}\tau_0}$ which seems to be compatible with what we observe. At large $\tau_0$, $\delta p / p$ should be constant and again this seems to be compatible with what we observe.
In Figure \ref{fig:errseed2} we plot the mean of the standard deviations as a function of $\tau_0$ for the edge-on view. Note that the scatter on $\delta_p$ is higher at larger $\tau_0$, which is simply due to the error bars being dependent on phase as we can see in the bottom graph of Figure \ref{fig:diffSeed}. However the mean standard deviation on the points still seems to rise as we go higher.
	  
\section{Comparison with the Analytical Results} 
\label{sec:comp}
\subsection{Optically Thin}
In \cite{Ignace2015}, we presented analytical calculations for the periodic polarization variability from CIRs embedded in an otherwise spherical wind for the optically thin case. In our second paper \citep{St-Louis2018} we applied a similar approach to optically thick winds. For our MCRT calculations, in order to differentiate between the effects of the CIRs and that of the wind, we proceeded in two steps. First we only included CIRs by imposing that the density in the grid cells not associated with them was 0. As a second step, we used a non-zero density for the spherical wind and added the density to that of the CIRs. 

\begin{figure*}
		\includegraphics[width=\textwidth]{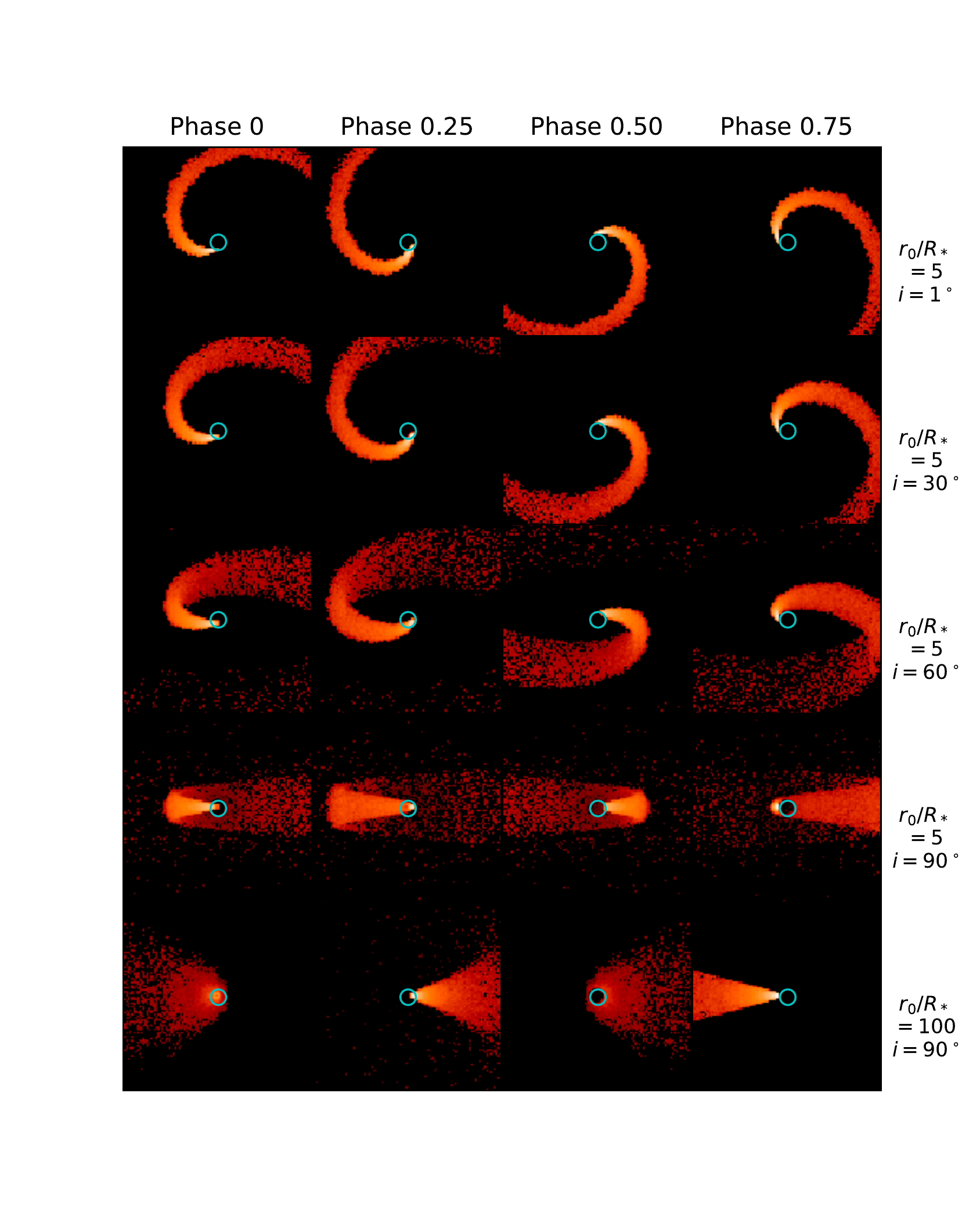}
 		\caption{Polarisation intensity images for one CIR placed at the stellar equator with $\tau _0$= 0.03. Each column corresponds to a different rotation phase, with phase 0 being the CIR footprint facing the observer. The first four rows are for $r_0/R_*$=5 with the stellar inclination varying from $i=$1$^\circ$ to $i=$90$^\circ$ and the last $r_0/R_*$=100. The blue circle shows the position of the stellar disk.}
 		\label{fig:polmaps}
	\end{figure*}

In Figure \ref{fig:polmaps}, we present a series of polarisation intensity images ($P=\sqrt{Q^2+U^2}$) for the case of an equatorial CIR only (no wind) and for a value of $\tau _0$= 0.03, which corresponds to the limit of an optically thin case. The four columns correspond to different rotational phases. Phase 0 is for the base of the CIR facing the observer and phase 0.5 is for when it is behind the star.  The first four rows are for a CIR with a modest winding radius of $r_0/R_*$=5 and for an inclination of the stellar axis of respectively $i=$1$^\circ$ (nearly pole-on), $i=$30$^\circ$, $i=$60$^\circ$ and $i=$90$^\circ$ (edge-on).  The last row present the edge-on view of an almost straight CIR ($r_0/R_*$=100). The polarisation intensity images are proportional to the density of the gas and therefore in addition to showing the distribution of the polarization, they also provide a map of the density structure of the CIR. We have added a blue circle to the maps to indicate the size of the stellar disk. 

\subsubsection{CIRs only}
\label{ssec:CIRonly}

In this section we compare our Monte Carlo polarization calculations with the results from our analytical model, first in the optically thin case and then in the optically thick limit. Our goal is to confront both approaches to verify if they agree and to bring to light any differences there may be.

We first discuss our results when assuming that only the grid cells containing the CIR have a density incremented by equation \ref{eq:rho_CIR} and that all other cells have a nil density. Note that we retain the same contrast for the CIRs with the wind. This way we will be able to compare the cases with and without a spherical wind. In Figure \ref{fig:Rico1}, we show a comparison between analytical (solid curves) and MCRT (filled circles) calculations for one CIR with $r_0 = 5R_*$ placed at different latitudes ($\theta_{\rm CIR}$ from $20^\circ$ to $80^\circ$). In the top panels, we show $p$ normalized by $\tau_0$ as a function of phase and $q$ vs $u$ for an inclination of $30^\circ$ and in the bottom panels, the same plots for an inclination of $60^\circ$. 

	We note the close similarities between the analytical and the Monte Carlo model. The curve is either single or double peaked, depending if the CIR is viewed in a more stationary manner by a given viewpoint. However, there still are some small differences that are larger than the numerical error, discussed in Section \ref{ssec:err}. In general, the MCRT values are below the analytical ones with the largest deviations at phases near 0.5 when the CIR is located behind the stellar axis. Strangely, there does not seem to be a coherent pattern in the deviations as a function of $\theta_{\rm CIR}$ with the best agreement for $\theta_{\rm CIR} = 20^\circ$ and $\theta{CIR} = 80^\circ$ and the worst for $\theta{CIR} = 40^\circ$. This behaviour is most likely specific to this particular viewing configuration.

\subsubsection{CIR with Wind}

\begin{figure}
		\includegraphics[width=\columnwidth]{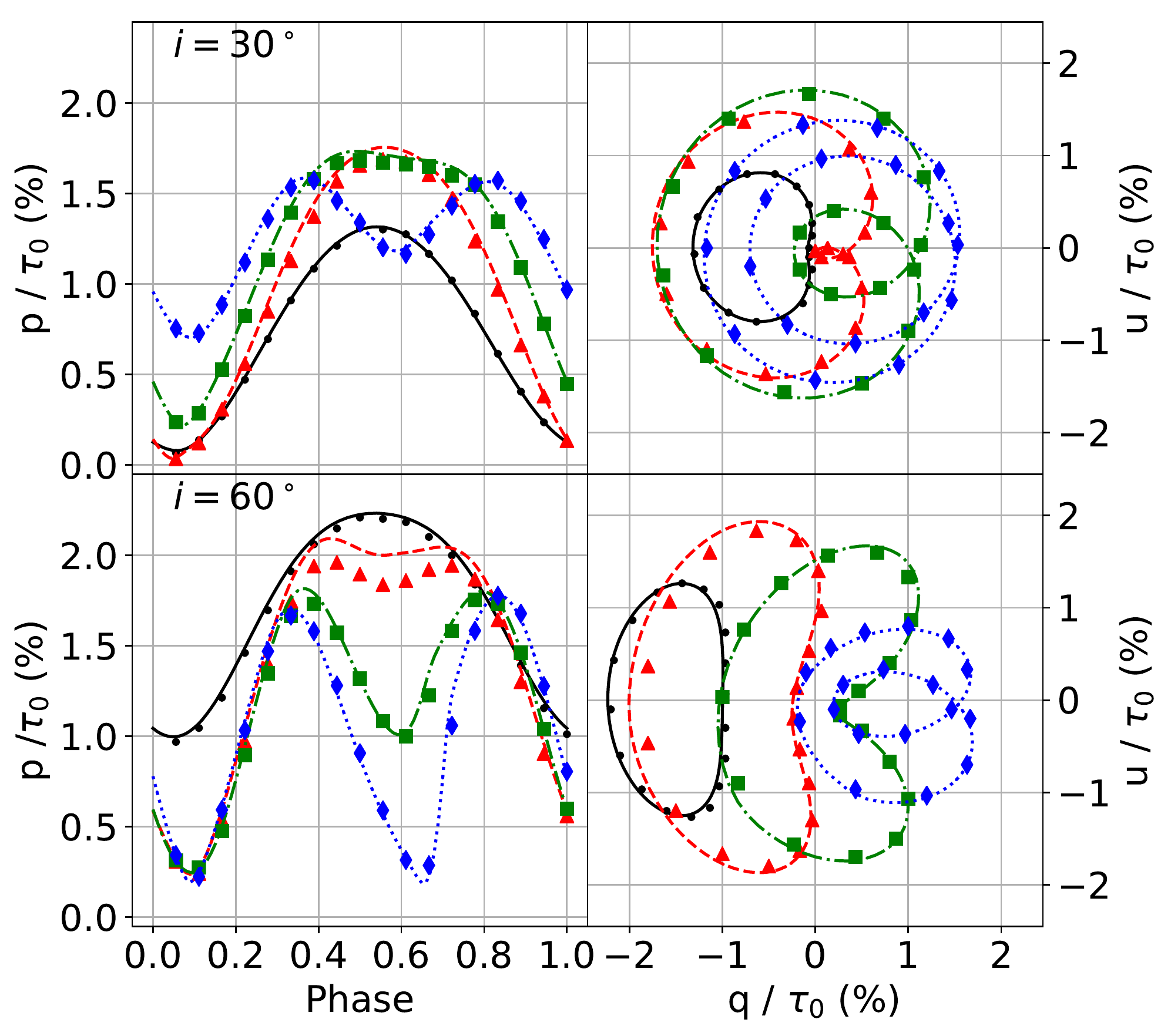}
 		\caption{Comparison between our MCRT linear polarization values (shapes) and those from the analytical model (lines) when only a CIR with $r_0 = 5R_*$ is included, meaning that the wind density is set to 0. We vary the CIR latitude $\theta_{\rm CIR}$ as we did in \citep{Ignace2015} and present results for an inclination of $i = 30^\circ$ for the top row and $i = 60^\circ$ for the bottom row. Calculations are carried out for $\tau_0 = 0.03$.  Filled circles (black) are for $\theta _{CIR}$=20$^\circ$, triangles (red) for $\theta _{CIR}$=40$^\circ$, squares (green) for $\theta _{CIR}$=60$^\circ$ and diamonds for $\theta _{CIR}$=80$^\circ$.}
 		\label{fig:Rico1}
	\end{figure}

\begin{figure}
		\includegraphics[width=\columnwidth]{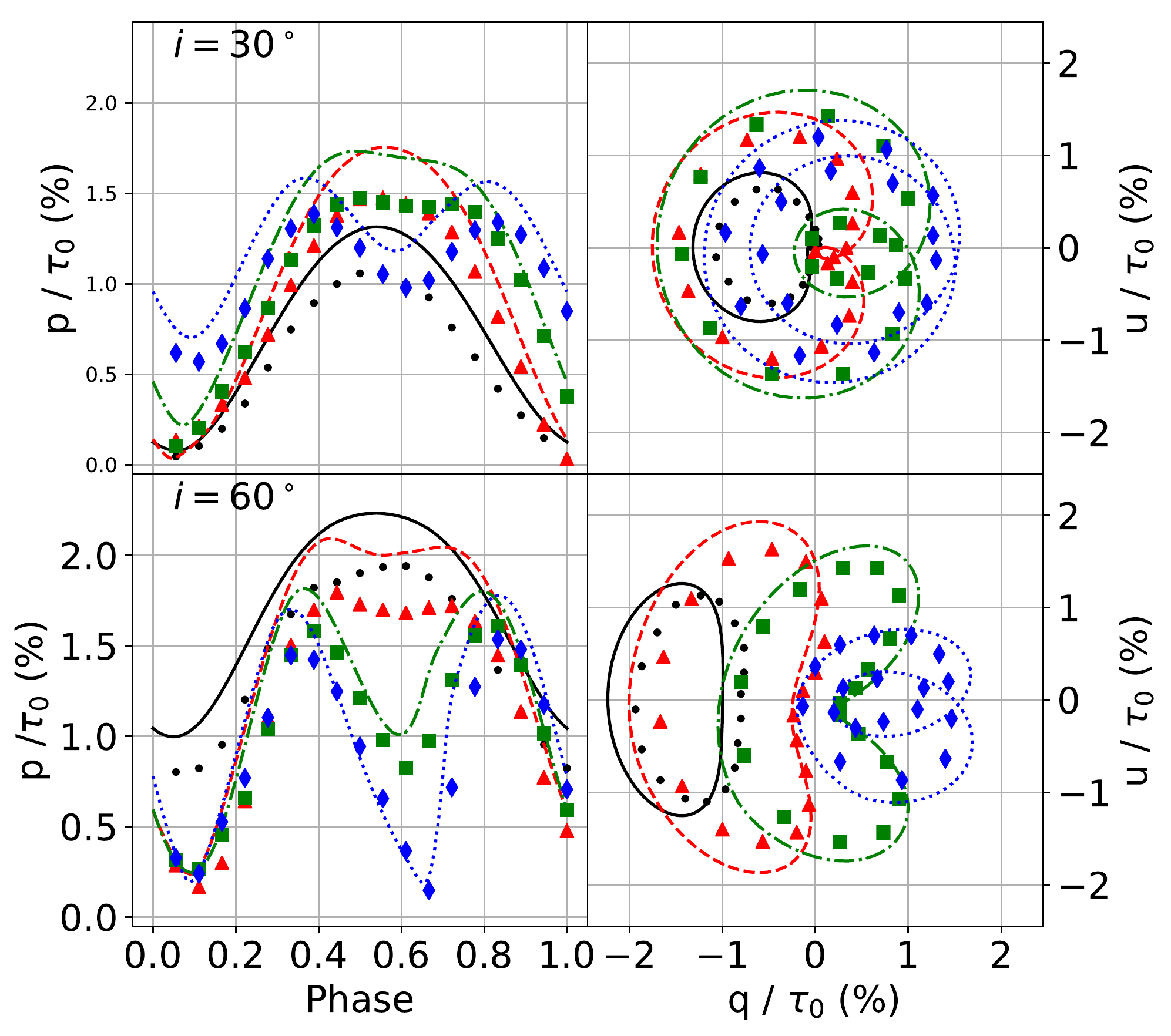}
 		\caption{Comparison between our MCRT linear polarization values (shapes) and those from the analytical model (lines) when both the CIR with $r_0 = 5R_*$ and the wind are included. We vary the CIR latitude $\theta_{\rm CIR}$ as we did in \citep{Ignace2015} and we present results for an inclination of $i = 30^\circ$ for the top row and $i = 60^\circ$ for the bottom row. Calculations are carried out for $\tau_0 = 0.03$. Symbols and colours as in Figure \ref{fig:Rico1}.}
 		\label{fig:Rico2}
	\end{figure}

Even though the net contribution from the spherical wind should in principle be zero, we have carried out the Monte Carlo simulations with a non-zero density for the wind. We present our results in Figure \ref{fig:Rico2}, superimposed on the same analytical curves as in Figure \ref{fig:Rico1}. One can see immediately that the differences are much more pronounced than for the case without a wind. Although the general form of the curves are the same, the polarization is attenuated for all phases and maybe even slightly shifted (see for example $i=60^\circ$, $\theta_{\rm CIR}=40^\circ$). Once again the largest differences are for phases near 0.5.

\subsubsection{What Causes the Differences?}
Part of the difference might be explained in the way our polarization values are normalized in equation \ref{eq:StNorm}. For the analytical calculations, since the scattering in the envelope is expected to be small compared to the direct star light, the total intensity was assumed to be that coming directly from the star, i.e. $I = I_*$. For the Monte Carlo calculations, all scattering contribution as well as the pre and post scattering attenuation are included by default since each photon run in the simulation contributes to $I$. In this interpretation, the differences are so much larger when we include the spherical wind because there are simply many more scatterings.
	
\begin{figure}
		\includegraphics[width=\columnwidth]{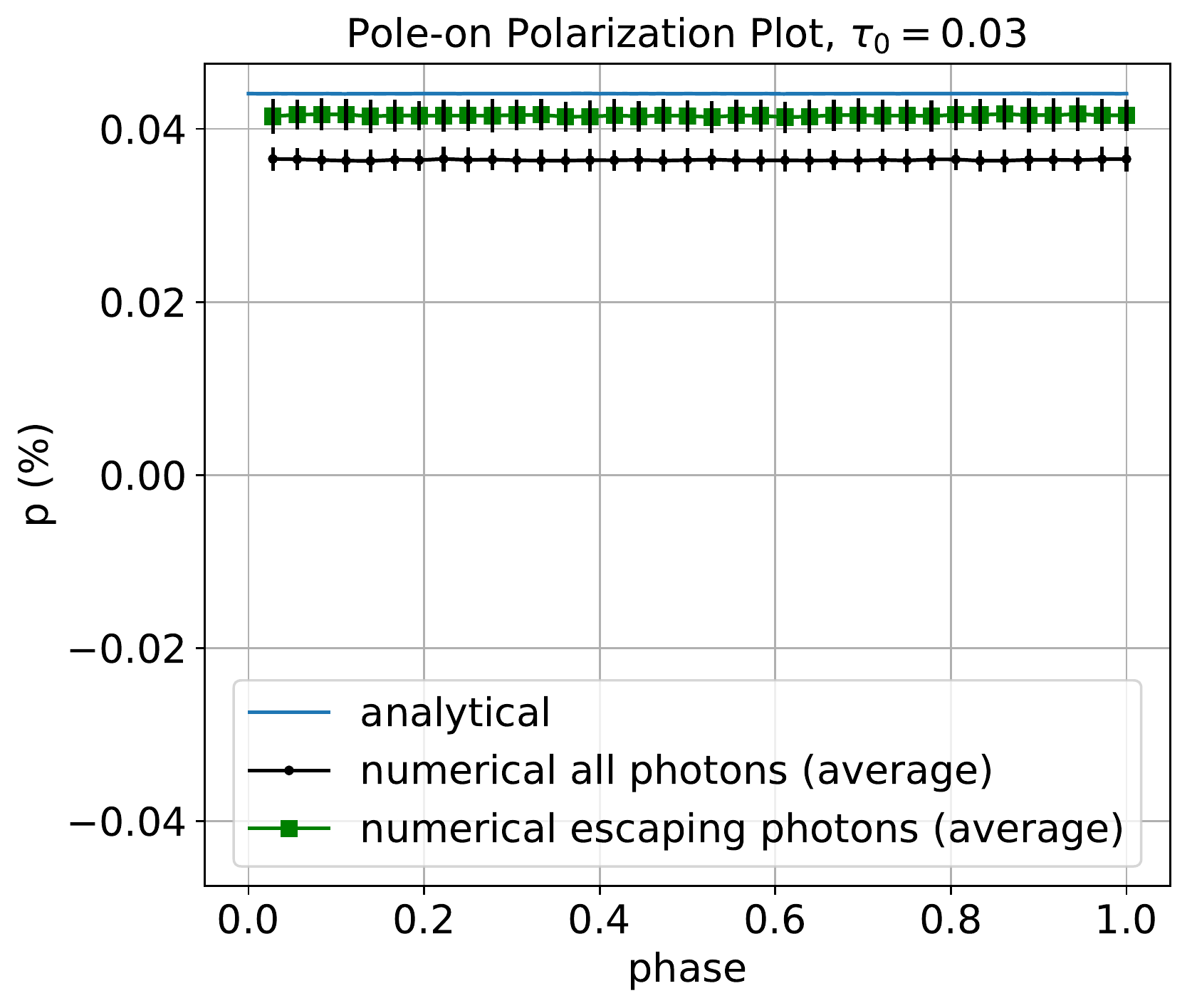}
 		\caption{Comparison between a polarization normalization by stellar escaping photons only and a normalization by all photons, for a wind with $\tau_0 = 0.03$ and a CIR with $r_0/R_* = 5$ in the pole-on view. Note that we still find a small difference with the analytical results, even with the different normalization. However the resulting polarization curve is much closer to the analytical results.}
 		\label{fig:stelphot}
	\end{figure}
	
To attempt to verify this hypothesis we ran simulations where the polarization bins were normalized by the flux of the escaping stellar photons $I_*$ only, instead of the total number of photons. In Figure \ref{fig:stelphot}, we compare the analytical values for a pole-on view of a equatorial CIR with $r_0 = 5R_*$ (blue curve) embedded in an optically thin wind with $\tau_0=0.03$ as well as the Monte Carlo results when our polarization values are normalized using all photons (black curve) or only the photons escaping directly from the star (green curve). Although the difference isn't compensated for completely, the resulting polarization is definitely much closer to the analytical results, with a deviation of only about 7\% between the two curves, as opposed to a difference of about 20\% between our original result and the analytical calculation. Note that this simple test does not remove the pre and post scattering attenuation terms which still contributes to the Monte Carlo value but obviously not to the analytical one.
	
\subsection{Optically Thick}
\begin{figure}
		\includegraphics[width=\columnwidth]{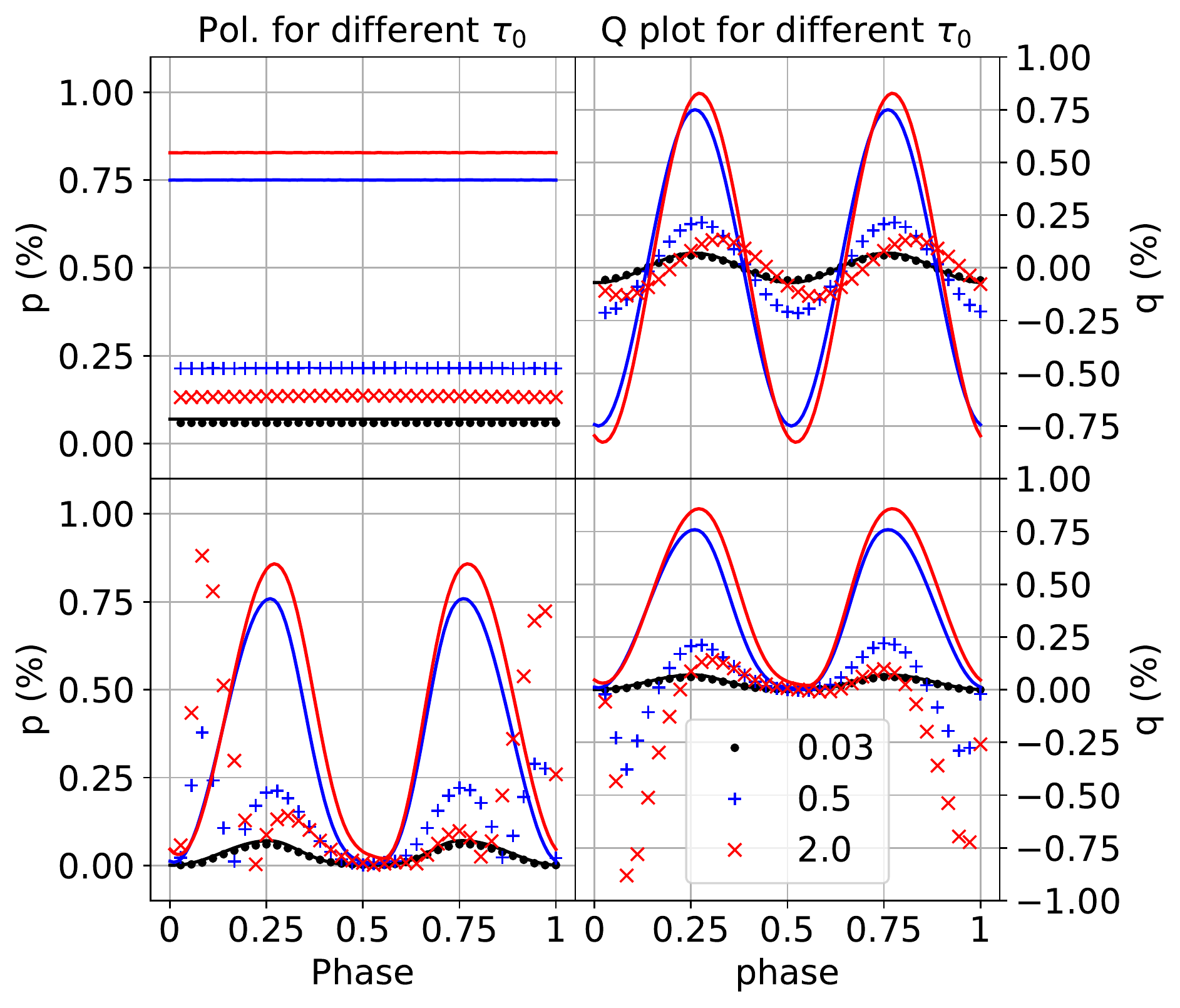}
 		\caption{Comparison between our MCRT linear polarization values (shapes) and those from the modified analytical solution (lines) for a wind containing a straight CIR with $r_0 = 100R_*$ at different $\tau_0$. Top row has a pole-on view $i = 1^\circ$ and the bottom row has an edge-on view, $i = 90^\circ$.}
 		\label{fig:Nicole1}
	\end{figure}

In this section we compare our numerical calculations with the analytical ones presented in \cite{St-Louis2018}, which treats optically thick winds in an approximate way using a "core-halo" approach. Since we already noted significant differences between the MCRT and the optically thin analytical calculations, we elected to adopt the simplest possible configuration. Therefore the calculations presented in this section are for an essentially straight CIR ($r_0 = 100R_*$) at the equator and we consider only the pole-on and edge-on views. In Figure \ref{fig:Nicole1} we compare analytical (solid curves) and MCRT (filled circles) results for three different values of $\tau_0$ : for an optically thin wind ($\tau_0 = 0.03$), for a moderate optical depth ($\tau_0 = 0.5$) and a strongly optically thick wind ($\tau_0 = 2.0$). The top row shows the total linear polarization $p$ (left), and Stokes parameter $q$ (right) as a function of phase for the pole-on view while the bottom row shows these same parameters for an edge-on view.
For the optically thin calculations (black curve and points), the difference between the MCRT and analytical points are the same order of magnitude as those shown in Figure \ref{fig:Rico1} and \ref{fig:Rico2}. For the pole-on view, for example, the difference in $p$ is around $0.008\%$. Note that unlike in the optically thin case, here we do not normalize our polarization values by $\tau_0$ because in the optically thick cases, the polarization does not scale linearly with $\tau_0$. as can readily be seen, the differences between the analytical and MCRT models are considerable for both optically thick cases. 

For the pole-on view, the amplitude of the polarization curves are wildly different, with the $\tau_0 = 0.5$ curve (blue) varying by about 0.55\% for $p$ and and the $\tau_0 = 2.0$ curve (red) varying by about 0.64\% for $p$. The red curve is also attenuated compared to the blue curve in the statistical approach, while the attenuation for the analytical approach has not happened yet. The curve shift in the statistical approach is also much more pronounced than it is in the analytical approach, varying by an amount of almost 0.1 phase between the blue and red curves, compared to the 0.01 change in phase that the analytical model has.

	In general, as the wind becomes increasingly optically thick, the differences between the MCRT and analytical models become increasingly large. For the pole-on view, for example, the difference is now $\Delta p = 0.5\%$ for $\tau_0 = 0.5$ and $\Delta p = 0.7\%$ for $\tau_0=2.0$. These are extremely large differences.
	
	The $q$ versus phase curve for the pole-on view presents an additional intriguing characteristic. As $\tau_0$ becomes larger, the maximum of the curve gradually shifts from the value of $\phi=0.25$ expected analytically to phases that are increasingly larger.
	
	For the edge-on case, the $q$ curve also presents this shift in its maximum but in addition, two dips to negative $q$ values appear on either side of phase 0 (CIR between the star and the observer). The $u$ curve (not shown here) shows values all close to 0, as expected, as the polarization vector is horizontal on the plane of the sky. These two dips are not predicted by our analytical model and makes $p$ curves ($\sqrt{q^2+u^2}$) complex looking. 

\subsection{Interpretation}\label{ssec:Interp}
The behaviour exhibited in these optically thick MCRT simulations are complex and the difference with the analytical calculations are large. In this section, we will present our interpretation of these results for both the pole-on and the edge-on views.
\subsubsection{Pole-on view}\label{sssec:PV}
\begin{figure}
		\includegraphics[width=\columnwidth]{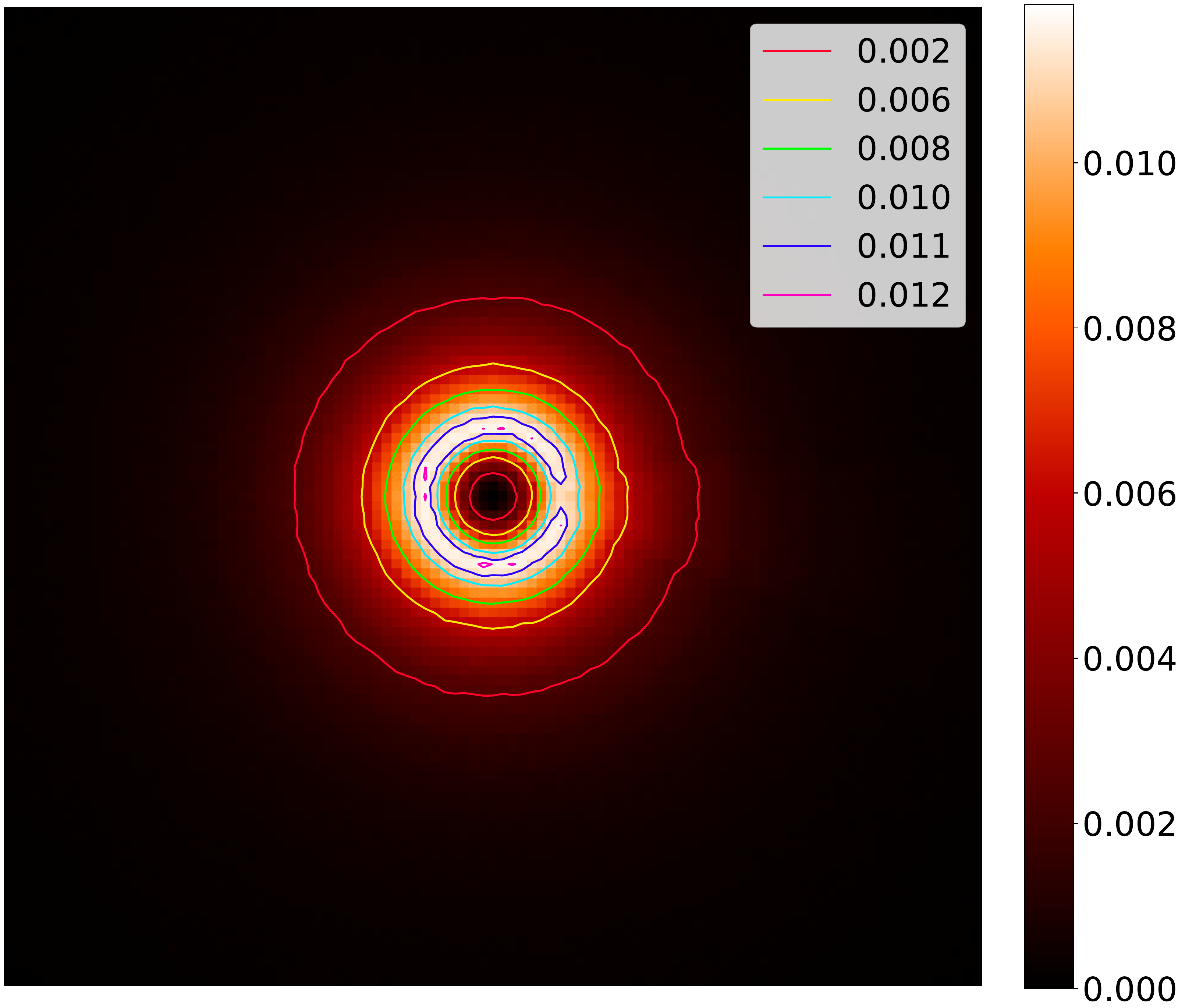}
		\caption{$24R_*$ by $24R_*$ linear polarization map (p) observed from the pole from a star with a spherical wind containing an essentially straight CIR ($r_0 = 100R_*$) for $\tau_0 = 2.0$, at phase $0.25$. Isophotes have been drawn, representing surfaces of constant polarization.}
 		\label{fig:Frame1}
\end{figure}

Figure \ref{fig:Frame1} shows a linear polarization image of a spherical wind and CIR at phase $0.25$ viewed from the pole. Superposed on the images are lines of various colours corresponding to isocontours on a linear scale. Moving radially from the center of the star, the polarization rises, reaches a maximum (region in white) and decreases again. This is a well-known behaviour for extended atmospheres of early-type stars \citep[e.g.][]{Brown1977, Cassinelli1987}. As with the analytical model, we see that the CIR causes an excess in polarization throughout most of the wind, except in the interior near the polarization. For the analytical calculations and our optically thin model this produces a double-wave $q$ curve with maxima at positive values at phases 0.25 and 0.75 and minima at negative values at phases 0 and 0.5 (see Figure \ref{fig:Nicole1}). However, in the optically thick cases, there is a deficit in the region where the polarization peaks at the location of the CIR. This can readily be seen as a break in the dark blue isocontour. This deficit introduces an important new contribution to the polarization of the wind as it breaks the previously axisymmetric polarization of the wind. The resulting curve has maxima at positive $q$ values at phase 0 and 0.5 and minima at negative $q$ values at phases 0 and 0.25. Both contributions (CIR and deficit) therefore vary in anti-phase, which greatly reduces the amplitude of the resulting polarization.

	There is one final ingredient that explains the apparent gradual shift in the $q$ curve with increasing $\tau_0$ that is seen in Figure \ref{fig:Nicole1}. As can be seen in Figure \ref{fig:Frame1}, even though our chosen CIR is essentially straight ($r_0 = 100 R_*$), a slight curvature is still present. This can be readily seen by measuring the position of the center of the CIR on the yellow and red isophotes. While the center of the deficit in the wind at $r = 1.7R_*$ is on the horizontal, the center of the yellow isophote at the position of the CIR is clearly below and the center of the CIR on the red isophote is even lower. This will produce a $q$ polarization curve that is slightly shifted from the one from an optically thin wind. As the wind becomes increasingly thick, the CIR will emerge at higher and higher radii shifting the curve accordingly to brighter phases. When the two polarization contributions are added (deficit plus slightly curved CIR) we obtain a curve with a greatly reduced amplitude with maxima that gradually shift towards higher phases as the optical thickness of the wind increases, as seen in Figure \ref{fig:Nicole1}.
	
\subsubsection{Edge-on view}\label{sssec:EV}
	For the edge-on view, there is a extra level of complexity because there are now occultation effects. In the bottom right panel of Figure \ref{fig:Nicole1}, two dips to negative values can be seen in the $q$ polarization curve, on either side of phase 0 for optically thick calculations ($\tau_0=0.5$ and $\tau_0 = 2.0$). As the optical thickness of the wind increases, the dips become deeper and they gradually move away from phase 0. The $q$ values remain positive in the other parts of the curve. Our interpretation of these dips is that we are seeing the occultation of sections of the wind polarization by the dense CIR on either side of phase 0.

\begin{figure}
		\includegraphics[width=\columnwidth]{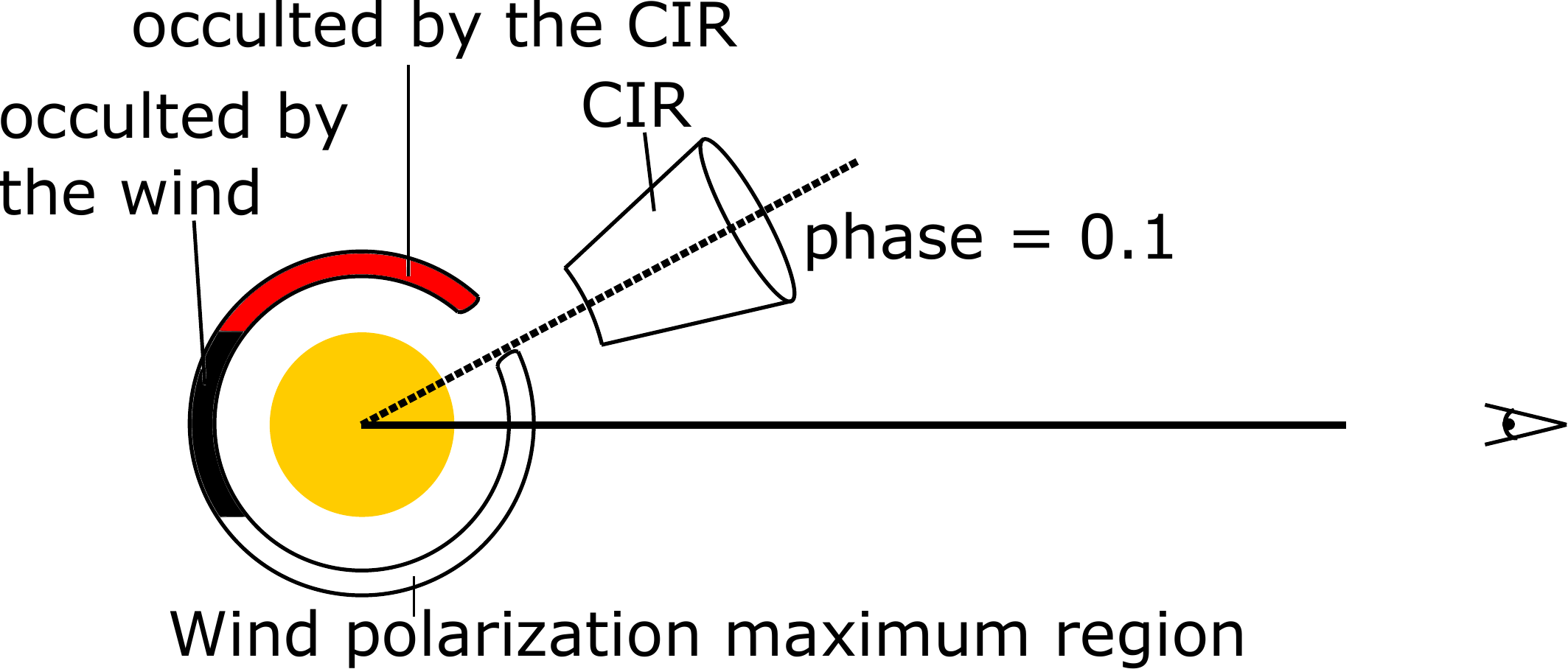}
		\caption{2D sketch of an observer (eye) viewing the star (centre circle) with a wind and CIR (2d cone) from an edge-on point of view at phase $\phi = 0.1$. The max polarization band in the wind is represented by the region between two disconnected and partially filled circles. The red or lighter band represents the region occulted by the CIR while the black or darker band represents the region occulted by the star. Note that the CIR isn't centered on the opening due to it's curvature, which is approximated in this sketch by a slightly displaced CIR.}
 		\label{fig:sketch}
\end{figure}

	In Figure \ref{fig:sketch} we present a 2D sketch of an edge-on view of the wind and CIR at a phase around $\phi = 0.1$. It can be seen that a large fraction of the wind polarization in the horizontal direction is occulted by the optically thick CIR. As a consequence, the balance between the horizontal and vertical components, previously leading to a nil polarization, is now broken and produces a net vertical polarization, i.e. negative $q$ values . This leads to the two dips in the $q$ curve. These two negative dips are superposed on a curve that is identical to the one that can be seen for the pole-on view and plotted in the top right panel of Figure \ref{fig:Nicole1}. indeed, the contributions from the spherical wind are the same whether they are viewed pole-on or edge-on. 
	
	The effects of the curvature are also visible in this edge-on view. At $\tau_0 = 0.5$, the CIR emerges closer to the star than at $\tau_0 = 2.0$. Therefore at $\phi=0$, the polarization reaches almost 0 for the $\tau_0 = 0.5$ case as the dominant part of the CIR is then symmetrical in our line of sight, which is not quite the case for the $\tau_0=2.0$ case. The curvature also manifest itself through the slight asymmetry of the two dips around phase 0, and the two peaks around phase 0.5. The slight curvature inward when the CIR is at phase 0.25 will scatter more photons into the line of sight than the outward curvature at 0.75 would. This also implies that at phase 0.1 more photons will be scattered out of the line of sight than at phase 0.9.
\section{Gaussian Spot Models}\label{sec:GauSpot}
In this section, we present MCRT calculations of the polarization, now including the total light intensity of a spherical wind and a CIR along with a stellar spot on the surface of the star at the footpoint of the CIR. Different spot models have been studied for a variety of star types, for example, in \cite{Al-Malki1992}, whose model generated small variations in polarization due to asymmetries in the photosphere. Here a gaussian spot model will be used.
\subsection{MCRT models including spots on the stellar surface}
\label{ssec:diffcomp}
Here we present results of MCRT models for a spherical wind spanning three values of $\tau_0$; and optically thin wind ($0.01$), a moderately thick wind ($0.1$) and a thick wind ($1.0$). We also include a CIR with $r_0 = 100R_*$ and a density contrast of $\eta=1$ at the stellar equator. Finally, we include a spot on the surface of the star, at the base of the CIR with the same angular extent as the CIR. We will vary the opening angle of the spot and CIR and the intensity of the spot with respect to the rest of the star.
\subsubsection{The effect on the Polarization curves}
\label{sssec:polcont}

	\begin{figure*}
		\includegraphics[width=0.8\textwidth]{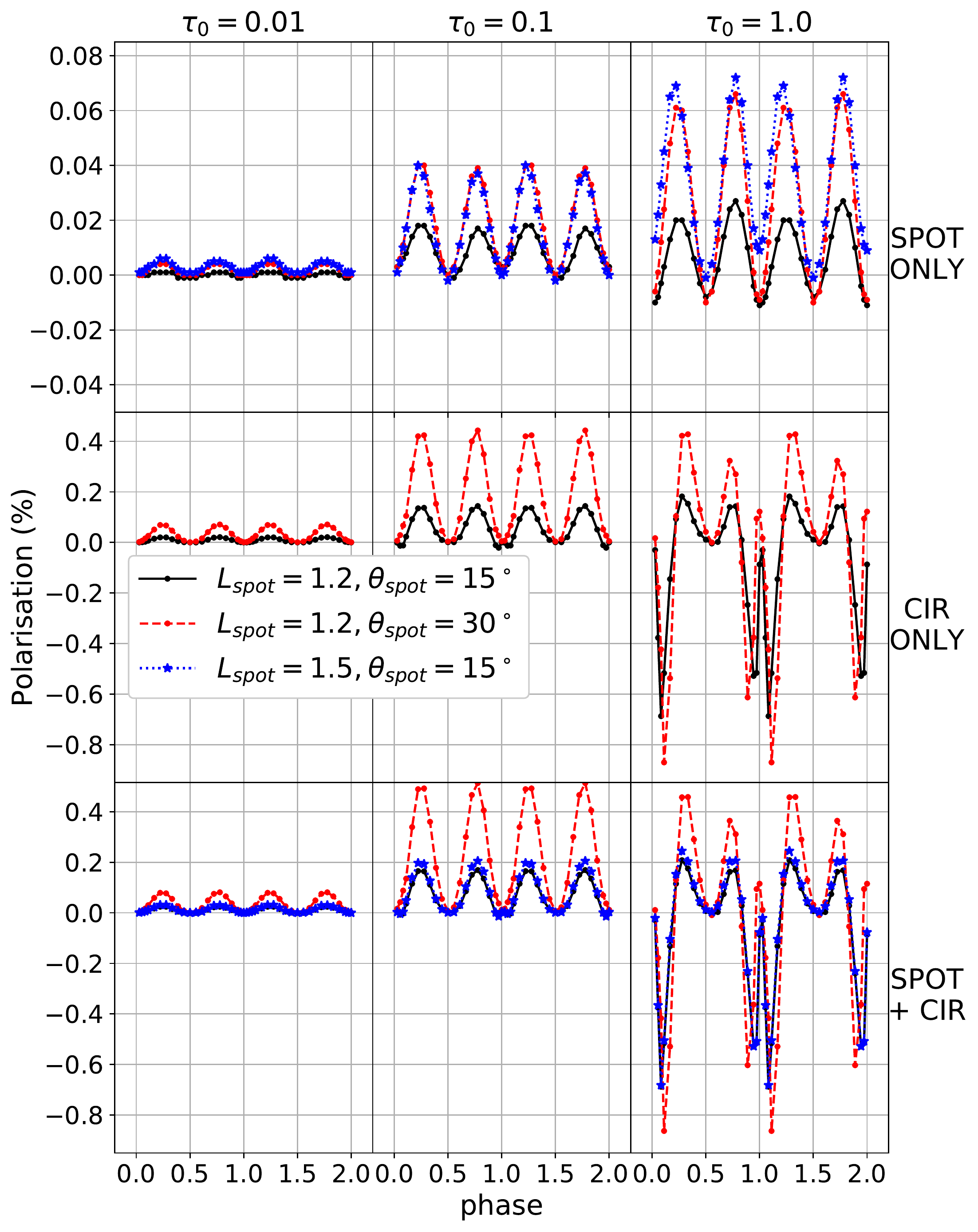}
 		\caption{Contributions to the $q$ polarization from the gaussian spot (top row), the CIR (middle row) and both combined (bottom row) for three different $\tau_0$ ; $\tau_0 = 0.01$ (left column), $\tau_0 = 0.1$ (center column) and $\tau_0 = 1.0$ (right column). Three different combinations of spot luminosity $L_{\rm spot}/L_{\rm phot}$ and spot opening angular radius $\beta$ were used. Note that the scale on the first row is different to the two other rows in order to see more clearly the amplitude of variation of intensity caused by the spot.}
 		\label{fig:gaussspot1}
	\end{figure*}
	In Figure \ref{fig:gaussspot1}, we present $q$ polarization curves for an edge-on view for three configurations. First, in the top for a spherical wind with only a spot at the surface of the star (no CIR). In the middle panels, we show a spherical wind with a CIR only (no spot). Finally in the bottom panels, we show results for the combination of a spot and a CIR. We also vary the spot and CIR parameters to get a better idea of their effect. In each plot, the black curve is a spot with a luminosity contrast of 1.2 and a half-opening angle of $15^\circ$. The blue curve is for the same opening angle but a luminosity contrast of 1.5. Finally, the red curve is for a spot with a 1.2 luminosity contrast but for a wider spot with $\beta = 30^\circ$.
	
		The effects on the $q$ polarization curves of the CIR only are as discussed in the previous section. Here in addition, we can see that increasing the opening angle increases the amplitude of the curve and the depth of the eclipses of the wind and can be explained within the framework of our interpretation.
		
		The effects on the polarization of a spot are illustrated in the top panels. First note that the amplitude of polarization is a factor of $\sim10$ smaller than in the case of a CIR only. Second, as expected, for an optically thin or modestly thick wind, the effect of increasing the brightness ratio of the opening angle is to increase the amplitude of the curve. This curve has two maxima per cycle, one at 0.25 and the other at 0.75 when the scattering angle is $90^\circ$ and two minima at $q=0$ when the spot is in the line of sight of the observer at phase 0 (forward scattering) or behind the star (occulted). The behaviour for the optically thick wind seems more complex when the spot is in front of the star at $\phi=0$, the $q$ polarization can either be positive if its contrast is higher (1.5) or negative if it is lower (1.2). When the spot is behind the star at $\phi=0.5$, the $q$ polarization is either 0 for a brighter spot (1.5) or negative for a lower luminosity contrast (1.2). This can be accounted for mainly by numerical noise, as the errors on the polarization values at $\tau_0=1$ in Figure \ref{fig:difftau} are quite large ($\sim0.03$) at this scale. As for the CIR, the amplitudes for the optically thick case at phase 0.25 and 0.75 are not quite equal. This can most likely be explained by the fact that the CIR is slightly curved, even for $r_0/R_* = 100$ and the leading and tailing edges then cause an asymmetry in the polarization (see Section \ref{sssec:EV}).
		
		In the bottom panels, we present the combined effects of the spot and CIR. the most important conclusion is that the effect of the spot on the polarization is similar in nature as that of the CIR (excluding the eclipse effects) but that they are of much smaller amplitude. Therefore, they do not affect significantly the shape of the polarization curves.
\subsubsection{The effect on the light-curves}
\label{sssec:intcont}
	\begin{figure*}
		\includegraphics[width=0.8\textwidth]{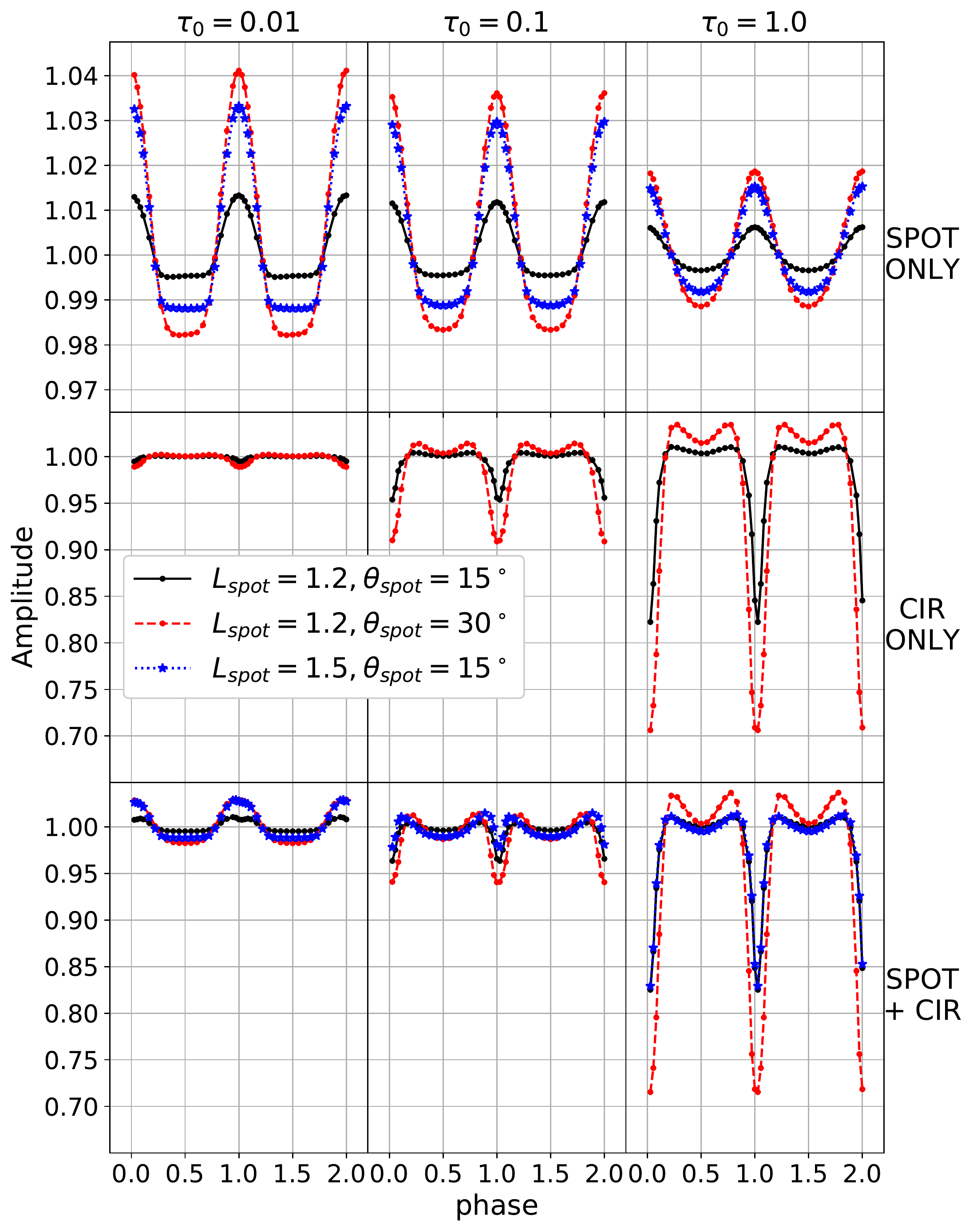}
 		\caption{Contributions to the intensity from the gaussian spot (top row), the CIR (middle row) and both combined (bottom row) for three different $\tau_0$ ; $\tau_0 = 0.01$ (left column), $\tau_0 = 0.1$ (center column) and $\tau_0 = 1.0$ (right column). Three different combinations of spot luminosity $L_{\rm spot}/L_{\rm phot}$ and spot opening angular radius $\beta$ were used. Note that the scale on the first row is different to the two other rows in order to see more clearly the amplitude of variation of intensity caused by the spot.}
 		\label{fig:gaussspot2}
	\end{figure*}
	
	In Figure \ref{fig:gaussspot2}, we present light-curves associated with the polarization curves presented in Figure \ref{fig:gaussspot1}. For the spot only, the curves are very much as expected with an increasing amplitude, when the spot is brighter and when it has a bigger surface. As the wind becomes thicker, the amplitude becomes smaller and smaller and the eclipse become less sharp. This is because as $\tau_0$ increases, light from the spot is diffused outward, making the spot larger and blurrier. For the CIR only, the behaviour is also as expected. When the CIR is in front at phase 0, it eclipses part of the star creating a dip. Of course, if the CIR is wider, the eclipse is also wider but also deeper. At phases 0.25 and 0.75, it scatters the light into the line of sight, generating excess light. For a wider CIR, these excesses are stronger. Finally, when the CIR is behind the star, it is totally invisible and the relative flux is unaffected ( = 1.0) for thin and moderately thick winds. For thick winds, some flux seems to reach the observer ( > 1.0) indicating that when it emerges, the CIR is slightly larger than the stellar photosphere. As for the relative amplitudes between the effects of the spot and that of the CIR, when the wind is thin (0.01) the spot dominates, but when the wind is thick (1.0) the CIR dominates. This is true even for moderately thick winds (0.1).

\subsection{MCRT CIR polarization curves for a range of densities}
\begin{figure}
		\includegraphics[width=\columnwidth]{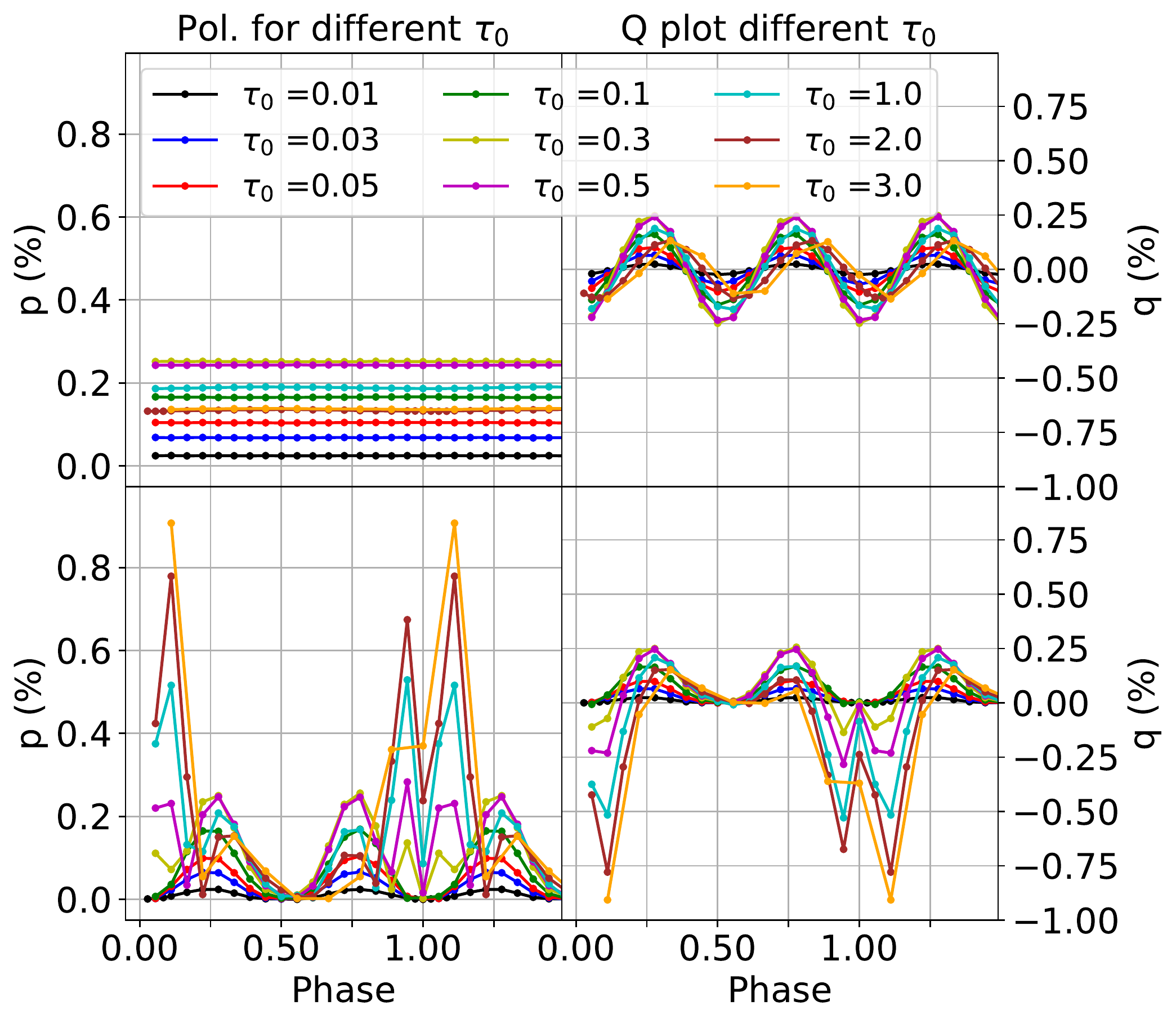}
 		\caption{Polarization values $p$ and $q$ as a function of phase of a spherical wind with a single essentially straight CIR ($r_0 = 100 R_*$) located at the equator for different values of $\tau_0$ from $0.01$ to $3.0$, with spot included for a pole-on view $i = 1^\circ$ (top row) and an edge-on view $i = 90^\circ$ (bottom row). Note that our $\tau_0 = 3.0$ simulation has half the amount of phase points, making the curve a bit less resolved. }
 		\label{fig:difftau}
	\end{figure}

In Figure \ref{fig:difftau}, we present polarization curves for a wind with an essentially straight equatorial CIR ($r_0/R_*=100$) in a pole-on (top row) and edge-on (bottom row) view for $p$ (left column) and $q$ (right column) as a function of phase for different values of $\tau_0$. For these models, we have also added a gaussian spot with $L_{\rm spot}/L_{\rm phot}=1.2$. Here 1.5 phase cycles are shown more clearly the shape of the curve. For the pole-on view, we can see that the total linear polarization, $p$, increases with $\tau_0$ until it reaches a maximum value of $\sim0.25\%$ at $\tau_0=0.3-0.5$. Above this value, increasing $\tau_0$ gradually decreases the value of $p$ until it reaches a value of $\sim0.15$ at $\tau_0=2.0$. Our calculation at $\tau_0=3.0$ gives a very similar polarization value. These effects can also be seen in the amplitude of the $q$ curves shown in the top right panel (the $u$ curves are in anti-phase with the $q$ curves). In addition to these variations in the amplitude of the $q$ curve with $\tau_0$, we can also see the gradual shift in the maxima of the curves, already described in Section \ref{sssec:PV}. This shift begins to become significant after $\tau_0=0.5$, approximately when the maximum in p is reached. This is consistent with our interpretation that at a certain value of $\tau_0$ ($0.3-0.5$) the optical depth in the CIR becomes important enough to break the symmetry of the wind polarization, hereby generating a new linear polarization source that varies in anti-phase with the polarization curve generated by the CIR itself. As $\tau_0$ increases, the CIR emerges at larger and larger distances from the star and because even with $r_0/R_*=100$ it still presents a slight curvature, the polarization curve from the CIR becomes gradually shifted to higher phases as $\tau_0$ increases.

	For the edge-on view, we can see the gradual appearance of the double dips caused by the eclipse of the wind by the CIR on either side of phase 0, also starting around $\tau_0=0.3$. These dips become deeper and wider as $\tau_0$ increases as the wind polarization becomes larger and the CIR occults a larger and larger fraction of the wind polarization.

One interesting thing to note here is that, in the edge-on view, we can see at what $\tau_0$ the two peaks in the polarization curve around phase 0 start appearing, in this case around $\tau_0 = 0.3$. With increasing density the peaks become higher.

As for the pole-on view, we can see that at in between $\tau_0 = 0.3$ and $\tau_0 = 0.5$, the polarization peaks reach a maximum and start decreasing. There also seems to be slight phase shifting in $q$ becoming most noticeable around $\tau_0 = 1.0$. Note that the decrease in polarization seems to have stopped in between $\tau_0 = 2.0$ and $\tau_0 = 3.0$, since the amplitude has stayed the same, however the shifting in $q$ still continues. 

In view of the fact that the polarization curves we obtain using MCRT for CIRs in a spherical wind are quite different from those obtained using analytical models, the fits of the observations of the WR star WR6 presented in \cite{St-Louis2018} need to be re-done.  The fact that our MCRT curves have a much smaller amplitude and, depending on the optical depth of the wind, have peaks that are shifted compared to those obtained with the analytical models will certainly result in different output parameters, such as the density contrast or the opening angle of the CIR and perhaps even in a different orientation of the stellar axis with respect to our line-of-sight.

However, performing such new fits is beyond the scope of this paper. First, we will need to calculate a grid of MCRT models by varying the many parameters of our model. Calculations, particularly at higher values of $\tau _0$  are very expensive in computing time.  Then, we would need to adjust the model curves to the observations using a robust fitting method such as, for example, a Monte Carlo Markov Chain technique such as used in the {\em emcee} package.  We intend to pursue such fits in an upcoming paper.

\subsection{MCRT Light Curves for a range of densities}
\label{sec:lc}

Our MCRT calculations also include monochromatic light curves for scattering of star light from both the wind and CIR. Total intensities include light from the star and scattered light, $I=1$ corresponding to the intensity from stellar light only. In this section, we present results showing how these vary for various wind densities.
	
\begin{figure}
		\includegraphics[width=\columnwidth]{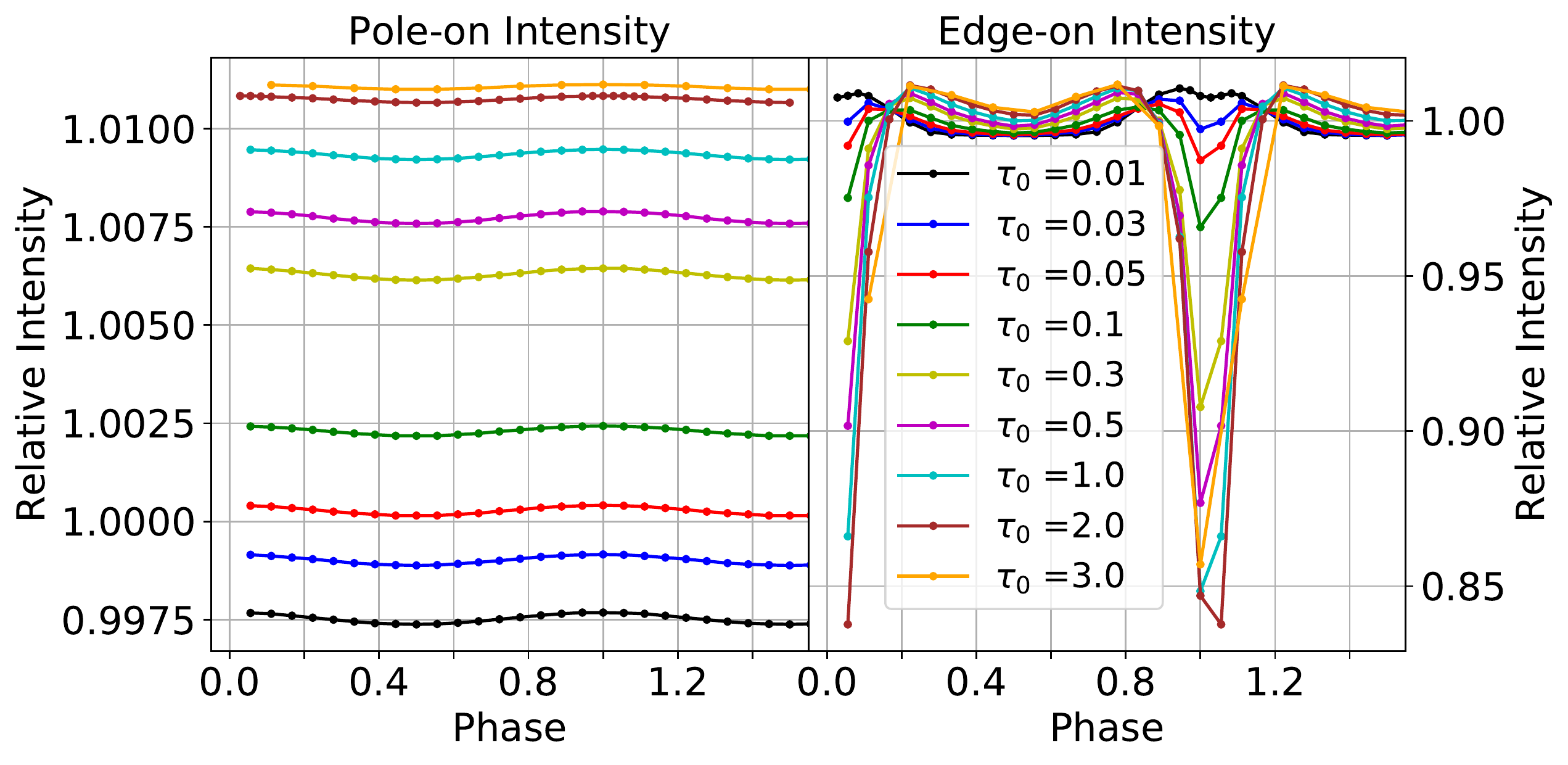}
 		\caption{Intensity values as a function of phase of a spherical wind with a single straight CIR ($r_0 = 100 R_*$) located at the equator for different values of $\tau_0$ from $0.01$ to $3.0$, with spot included for a pole-on view $i = 1^\circ$ (left) and an edge-on view $i = 90^\circ$ (right). Note the scale for the pole-on intensity is different from the edge-on intensity to highlight the difference between the different curves.}
 		\label{fig:lc100}
	\end{figure}
	
In Figure \ref{fig:lc100}, we illustrate how the light curves evolve as a function of $\tau_0$ for a spherical wind containing a single straight CIR ($r_0 = 100R_*$) at the equator with a spot on the surface of the star for pole-on (left) and edge-on (right) views. As expected, the pole-on case yields constant values of $I$ with an increase in amplitude for higher values of $\tau_0$. The slight systematic variations at this scale are due to the slight inclination ($\sim1^\circ$). The edge-on case shows curves with the same characteristics as those presented in Section \ref{sssec:intcont}, where the CIR contribution becomes gradually more important as $\tau_0$ increases. For low $\tau_0$ values, we can see a broad contribution from scattered light centered on phase 0 for this essentially straight CIR. As $\tau_0$ increases, this excess becomes more and more reduced by the more narrow dip generated by the eclipse of the wind of the star by the CIR. Around phases 0.25 and 0.75 however, as the CIR exits the line of sight of the star, we notice two bumps in the light curve, indicating an excess of photons scattered into the line of sight.

	\begin{figure}
		\includegraphics[width=\columnwidth]{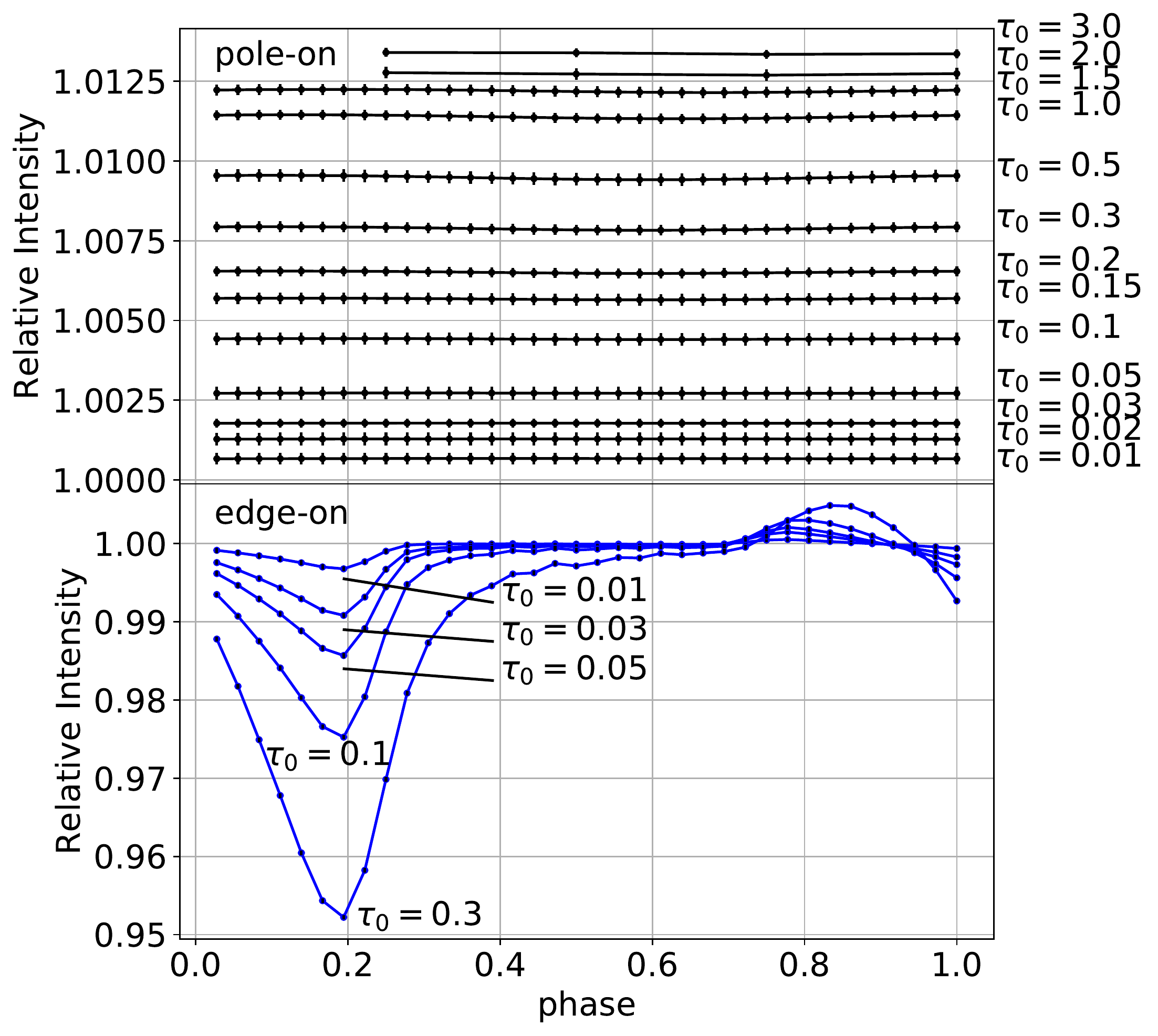}
 		\caption{Pole-on (top graph) and edge-on (bottom graph) intensity values as a function of phase for a spherical wind containing a single CIR with $r_0 = 5 R_*$ located at the equator for different values of $\tau_0$ from $0.01$ to $3.0$ for the pole-on view and from $0.01$ to $0.3$ for the edge-on view, with spot included. The error bars represent the standard deviation given by running the same simulation with 20 different seeds. }
 		\label{fig:lc5}
	\end{figure}
	
Figure \ref{fig:lc5} presents the light curves from a slightly curved CIR ($r_0 = 5R_*$) instead of a straight one, for different values of $\tau_0$ for a pole-on view (top) and edge-on view (bottom). For these curves, we have repeated the calculation 20 times and present the mean values on the figure. The error bars correspond to the standard deviation of these means. It can readily be seen that these intensity error bars are much smaller than their polarization counterparts in Figure \ref{fig:diffSeed}, as the errors are in general of the order of 0.0001.

\section{Conclusions}\label{sec:conc}
In this paper, we have shown that the results from our Monte Carlo statistical approach for treating CIRs differ significantly from the analytical models. While the Monte Carlo model fits relatively well with the optically thin results of \cite{Ignace2015}, only with some minor differences when we do not include the spherical wind with the CIR, the polarization becomes much more attenuated compared to the analytical model when we do include the wind in the Monte Carlo simulations. We interpret this as indicating that the scattered light and/or the pre and post scattering attenuation have a much more important impact than previously envisaged on the polarization with a decrease of about $\sim20\%$. When we compare our results with those of \cite{St-Louis2018} for the optically thick limit, the differences become even more important, as multiple scattering adds complexity to the polarization curves. First, the scattering of the photons by the CIR towards the line of site at phases 0.25 and 0.75 for an edge-on view is increasingly reduced as the optical depth becomes higher by the eclipse by the CIR of the polarized spherical wind on either side of phase $0$ (CIR in front). Secondly, because of multiple scattering, the optically thick CIR introduces a deficit in the region of maximum polarization, yielding a polarization contribution almost completely in anti-phase with the polarization generated by the CIR further out in the wind where the density is smaller. These two contributions in almost complete anti-phase greatly reduce the amplitude of the resulting polarization. Of course, the fits to observations presented in our previous paper using analytical curves need to be revisited and will certainly yield different stellar and CIR parameters.

	Adding spots on the surface of the star at the base of the CIRs has a small effect on the polarization curve, where a slight excess can be observed. However in the total light curves, three cases can be distinguished depending on the spot parameters and the optical thickness of the wind: The first is when the spot dominates in the optically thin limit, the second is when the CIR dominates in the strongly optically thick limit, and the last is when both contributions are significant, in the moderately optically thick limit.
	
	Although this statistical model presented in this paper is relatively simple, we believe it provides a base on which we will be able to build upon. In the future we plan to treat more complex wind and CIR geometries and kinematic structures such as those that result from hydrodynamical simulations.
	
\section*{Acknowledgements}
NSL acknowledges  financial support from the Natural Sciences and Engineering Research Council (NSERC) of Canada. RI acknowledges support by the National Science Foundation under Grant No. AST-1747658.   Computations were made on the supercomputer Briar\'ee from the Universit\'e de Montr\'eal, managed by Calcul Qu\'ebec and Compute Canada. The operation of this supercomputer is funded by the Canada Foundation for Innovation (CFI), the Minist\`ere de l'\'economie, de la science et de l'innovation du Qu\'ebec (MESI) and the Fonds de recherche du Qu\'ebec - Nature et technologies (FRQ-NT).




\bibliographystyle{mnras}
\bibliography{Carlos-Leblanc} 





\appendix




\bsp	
\label{lastpage}
\end{document}